\documentclass[]{emulateapj}
\usepackage{apjfonts}
\newcommand{\be}{\begin{equation}}
\newcommand{\ee}{\end{equation}}
\newcommand{\bea}{\begin{eqnarray}}
\newcommand{\eea}{\end{eqnarray}}

\newcommand{\Msun}{M_{\odot}}
\newcommand{\kms}{{\,{\mathrm{km}}\,{\mathrm{s}}^{-1}}}

\newcommand{\comment}[1]{}

\shortauthors{CONROY \& GUNN}
\shorttitle{SPS Model Calibration, Comparison, and Evaluation}
\begin{document}
\journalinfo{The Astrophysical Journal}
\submitted{Submitted to the Astrophysical Journal}

%----------------------------------------------------------------
\title{The propagation of uncertainties in stellar population
  synthesis modeling III: model calibration, comparison, and
  evaluation}
%----------------------------------------------------------------

\author{Charlie Conroy \& James E. Gunn}
\affil{Department of Astrophysical Sciences, Princeton
  University, Princeton, NJ 08544, USA}

\begin{abstract}

  Stellar population synthesis (SPS) provides the link between the
  stellar and dust content of galaxies and their observed spectral
  energy distributions.  In the present work we perform a
  comprehensive calibration of our own flexible SPS (FSPS) model
  against a suite of data.  These data include ultraviolet, optical,
  and near--IR photometry, surface brightness fluctuations, and
  integrated spectra of star clusters in the Magellanic Clouds (MCs),
  M87, M31, and the Milky Way (MW), and photometry and spectral
  indices of both quiescent and post--starburst galaxies at $z\sim0$.
  Several public SPS models are intercompared, including the models of
  Bruzual \& Charlot (BC03), Maraston (M05) and FSPS.  The relative
  strengths and weaknesses of these models are evaluated, with the
  following conclusions: 1) The FSPS and BC03 models compare favorably
  with MC data at all ages, whereas M05 colors are too red and the
  age--dependence is incorrect; 2) All models yield similar optical
  and near--IR colors for old metal--poor systems, and yet they all
  provide poor fits to the integrated $J-K$ and $V-K$ colors of both
  MW and M31 star clusters; 3) FSPS is able to fit all of the
  ultraviolet data because both the post--AGB and horizontal branch
  evolutionary phases are handled flexibly, while the BC03 and M05
  models fail in the far--UV, and both far and near--UV, respectively;
  4) All models predict $ugr$ colors too red, D$_n4000$ strengths too
  strong and H$\delta_A$ strengths too weak compared to massive red
  sequence galaxies, under the assumption that such galaxies are
  composed solely of old metal--rich stars; 5) FSPS and, to a lesser
  extent, BC03 can reproduce the optical and near--IR colors of
  post--starburst galaxies, while M05 cannot.  Reasons for these
  discrepancies are explored.  The failure at predicting the $ugr$
  colors, D$_n4000$, and H$\delta_A$ strengths can be explained by
  some combination of a minority population of metal--poor stars,
  young stars, blue straggler and/or blue horizontal branch stars, but
  not by appealing to inadequacies in either theoretical stellar
  atmospheres or canonical evolutionary phases (e.g., the main
  sequence turn--off).  The different model predictions in the
  near--IR for intermediate age systems are due to different
  treatments of the TP--AGB stellar evolutionary phase.  We emphasize
  that due to a lack of calibrating star cluster data in regions of
  the metallicity--age plane relevant for galaxies, all of these
  models continue to suffer from serious uncertainties that are
  difficult to quantify.

\end{abstract}


%-----------------------------------------------------------------

\section{Introduction}
\label{s:intro}

The spectral energy distribution (SED) of a galaxy contains a wealth
of information regarding its star formation history, dust content, and
chemical abundance pattern.  These properties provide essential clues
to the physical processes governing the formation and evolution of
galaxies from high redshift to the present.  It is therefore highly
desirable to have a robust method for extracting the physical
properties of galaxies from their SEDs.

The process of translating observed SEDs into physical properties is,
unfortunately, very challenging because it requires 1) an accurate
understanding of all phases of stellar evolution, 2) well--calibrated
stellar spectral libraries for converting stellar evolution
calculations into measurable fluxes, 3) an initial mass function
(IMF), specifying the weight given to each stellar mass, 4) detailed
knowledge of the star--dust geometry in conjunction with an
appropriate extinction curve; i.e., knowledge of the physical
conditions of the interstellar medium (ISM).  Each of these
requirements depend on chemical composition, further compounding the
problem.  Combining these ingredients in order to predict the spectrum
of a galaxy is known as stellar population synthesis (SPS), and has an
extensive history \citep[e.g.,][]{Tinsley76, Tinsley80, Bruzual83,
  Renzini86, Buzzoni89, Bruzual93, Worthey94, Maraston98, Leitherer99,
  Fioc97, Vazdekis99, Yi03, Bruzual03, Jimenez04, LeBorgne04,
  Maraston05, Schiavon07, Coelho07, Conroy09a, Molla09, Kotulla09}.

In the past two decades enormous progress has been made on each of the
requirements mentioned above.  Yet, substantial uncertainties remain.
A non--exhaustive list includes 1) the treatment of the core
convective boundary in main sequence stars (i.e., convective core
overshooting); 2) metallicity--dependent mass--loss along the red
giant branch (RGB) and, relatedly, the morphology of the horizontal
branch (HB); 3) the treatment of the thermally--pulsating asymptotic
giant branch (TP--AGB) phase; 4) the spectral libraries, especially
for M giants, TP--AGB stars, and stars at non--solar metallicities
\citep[e.g.,][]{Martins07}; 4) blue straggler (BS) stars and their
ubiquity \citep[e.g.,][]{Preston00, Li08a}; 5) the effects of
binarism, which might be especially relevant for massive star
evolution \citep[e.g.,][]{Eldridge08} and BS stars; 6) the importance
of rotation on massive star evolution \citep[e.g.,][]{Meynet00}; 7)
non--solar abundance ratios, which effect not only the stellar
evolution calculations but also the stellar spectra
\citep[e.g.,][]{Coelho07}; 8) the unknown dependence of the IMF on ISM
properties such as metallicity and pressure.  These uncertainties can
in many cases dramatically impact the ability to convert observables
into physical properties and vice--versa \citep[e.g.,][]{Charlot96a,
  Charlot96b, Yi03a, Gallart05, Maraston06, LeeHC07, Conroy09a, Muzzin09,
  Conroy09c}.

Thankfully, there exists a wide array of data capable of constraining
SPS models.  By far the most common type of object used for comparison
is the star cluster\footnote{Herein both open and globular clusters
  will be referred to as star clusters.}.  The approximately uniform
age and metallicity of the stars within star clusters and the lack of
internal reddening (except perhaps for very young clusters) affords
direct comparison between them and the most basic ingredients in SPS
--- the stellar evolution calculations and stellar spectral libraries.
Comparisons to star clusters are either made in CMD space or in an
integrated sense (i.e., the light from all the cluster stars are added
together).  The former technique provides a sensitive probe of the
main--sequence including the turn-off point, sub--giant and red giant
branches \citep[e.g.,][]{Worthey94, Bruzual03, An09}, while the latter
technique provides a more robust measure of the rarer brightest stars
that dominate the integrated light from the cluster.  It is the latter
method of comparison that is more relevant if the goal of SPS modeling
is to understand the integrated light from galaxies.  For this reason,
the integrated colors, surface brightness fluctuations (SBFs), and
spectra of star clusters have been used extensively to compare and
calibrate SPS models \citep[e.g.,][]{Bruzual03, Gonzalez04,
  Maraston05, Cohen07, Cordier07, Pessev08, Marigo08, Koleva08,
  LeeHC09b}.

There are two significant sources of concern when attempting to
calibrate SPS models with star cluster data.  The first concern is
that the brightest stars, which dominate the integrated light, are
rare, and thus stochastic effects must be carefully modeled.
Recently, this issue has been addressed observationally by stacking
clusters in bins of age in order to synthesize `superclusters' that
are relatively unaffected by stochastic effects \citep{Pessev06,
  Gonzalez04}.

The second concern is less tractable, even in principle, and arises
from the fact that the primary sources of star clusters for which ages
and metallicities can be reliably estimated are the Milky Way (MW),
M31, and the Large and Small Magellanic Clouds (LMC and SMC,
respectively).  This fact imposes severe restrictions on the region of
the age--metallicity parameter space that can be constrained with star
cluster data.  For example, one of the most important regions of this
parameter space for studying galaxies --- old and metal rich ---
contains very few star clusters (except in the bulge of the MW,
although high extinction diminishes the utility of star clusters
there).  In Local Group star clusters, a bin in age contains clusters
of a relatively narrow range in metallicity owing to the simple fact
that the metallicity of galaxies tends to increase with age.

Because of these issues, entire galaxies are often also used to assess
the accuracy of SPS models \citep[e.g.,][]{Worthey94, Bruzual03,
  Maraston06, Eminian08}.  Galaxies are not subject to the two
concerns mentioned above, but using them to constrain SPS models can
be very difficult because 1) galaxies contain stars of a range of ages
and metallicities; and 2) starlight in galaxies is attenuated by
interstellar dust.  Nevertheless, if subsamples of galaxies are
carefully constructed with known and regular properties, then galaxies
can be used to assess the reliability of SPS models.

The present work seeks to provide a comprehensive comparison between
SPS models and observed star clusters and galaxies.  Such a comparison
is timely because of an abundance of new, high--quality data,
including near--IR photometry and SBFs of star clusters in the LMC and
SMC, UV photometry of clusters in M31 and M87, CMDs and integrated
spectra of clusters in the MW, and optical and near--IR photometry and
spectral indices for large samples of low--redshift galaxies.  We will
consider not only our own SPS model \citep{Conroy09a} but also the
commonly used models of \citet[][BC03]{Bruzual03} and
\citet[][M05]{Maraston05}.  Such a detailed comparison between these
popular models has not been undertaken until now, and will thus
provide a much--needed evaluation of the relative strengths and
weaknesses of these models.  In addition, the flexible nature of our
own model will be exploited to explore the extent to which various
uncertain phases of stellar evolution, including thermally--pulsating
asymptotic giant branch stars (TP--AGB), blue HB stars, and post--AGB
stars, can be constrained by the data.  This task is critical if we
are to have confidence in the derived physical properties of galaxies,
such as stellar masses and star formation rates.

We proceed as follows. $\S$\ref{s:data} contains a detailed
description of the star cluster and galaxy data used to constrain the
models, which are themselves described in $\S$\ref{s:model}.  In
$\S$\ref{s:calib} we present an extensive comparison between models
and data.  A discussion and summary are provided in $\S$\ref{s:disc}
and $\S$\ref{s:sum}, respectively.  The zero points of the star
cluster UBVRIJHK magnitudes are in the Vega system; all other
magnitudes are quoted in the $AB$ system \citep{Oke83}.  Where
necessary, we adopt a flat $\Lambda$CDM cosmology with $(\Omega_m,
\Omega_\Lambda,h)=(0.26,0.74,0.72)$.

%-----------------------------------------------------------------

\section{Data Compilation}
\label{s:data}

This section describes in detail the sources and treatment of data
both for star clusters ($\S$\ref{s:scd}) and galaxies ($\S$\ref{s:gal}).

\subsection{Star cluster data}
\label{s:scd}

The calibration of SPS models against data is essential because of the
many uncertain model ingredients, including late stages of stellar
evolution and stellar spectral libraries.  As mentioned in the
introduction, the star cluster is the ideal observational datum for
SPS calibration because it is, at least approximately, a coeval set of
stars at a single metallicity.  In order to provide comprehensive
constraints on SPS models, one would like to have data on star
clusters spanning a wide range in age, metallicity, and wavelength.
Unfortunately, owing to the details of the chemical history of the MW
and its satellites, the assembly of such a dataset from nearby
galaxies, including our own, is challenging.

In the present work attention is focused on data from the Milky Way
(MW), M31, and the Magellanic Clouds (MCs).  The first two galaxies
are sources primarily of old, metal--poor clusters, while the MCs
provide clusters spanning a wider range of ages and metallicities.

Before describing the datasets in detail, it is worth emphasizing
those areas of parameter space that are not well--sampled and yet are
important for interpreting the integrated light from galaxies.
Massive star clusters ($M\gtrsim10^4\Msun$) at solar or super--solar
metallicities are not common in the Local Group.  Less massive star
clusters of such metallicities are more common, but stochastic effects
complicates model comparison.  The star clusters NGC 6791, NGC 188,
and the bulge star clusters NGC 6553 and NGC 6528 are notable
exceptions in that they are relatively old and metal--rich.  These
clusters will be used herein for calibration.

\subsubsection{The Milky Way}

Data on MW star clusters are taken primarily from the globular cluster
catalog of \citet{Harris96}.  This catalog provides $UBVRI$
photometry, [Fe/H] metallicities, and $E(B-V)$ reddening values for
150 clusters.  The photometry is corrected for Galactic extinction
using the reddening values provided in the catalog in conjunction with
a \citet{Cardelli89} extinction curve.  The \citet{Harris96} catalog
is complemented with $JHK$ photometry from the Two Micron All Sky
Survey \citep[2MASS;][]{Skrutskie06} for 106 clusters as compiled by
\citet{Cohen07}.  For most of the comparisons between models and data
the MW sample is restricted to clusters with $E(B-V)<0.2$ in order to
reduce the uncertainties associated with de--reddening.  When
presenting $V-K$ and $J-K$ colors, all clusters with $E(B-V)<0.4$ are
included since reddening corrections are smaller in the near--IR.

\citet{Schiavon05} have measured integrated, flux--calibrated, high
signal--to--noise spectra of 40 MW star clusters spanning a range of
metallicities and HB morphologies.  The spectra span the wavelength
range $3360<\lambda<6430$\AA\, at a FWHM resolution of 3.1\AA\, and
sampling of 1.0\AA.  These spectra have been de-reddened using the
\citet{Cardelli89} extinction curve with $E(B-V)$ values tabulated by
Schiavon et al.  Metallicities are also provided by Schiavon et
al. and will be used herein, except for 19 clusters that have updated
[Fe/H] measurements from \citet{Carretta09}.

\subsubsection{M31}

Photometry of star clusters in M31 is available in the Revised Bologna
Catalog \citep[RBC v3.5;][]{Galleti04}.  Available photometry includes
$UBVRI$ from a variety of sources, $JHK$ from 2MASS, and new
ultraviolet observations from the {\it Galaxy Evolution Explorer}
\citep[{\it GALEX};][]{Martin05} presented in \citet{Rey07}.
Spectroscopic metallicities of a subset of M31 star clusters are taken
from \citet{Barmby00} and \citet{Perrett02}.  Galactic $E(B-V)$
reddening estimates are derived from \citet{Schlegel98} via the
utilities in the \texttt{kcorrect v4.1.4} software package
\citep{Blanton07}.  Photometry is corrected for Galactic extinction
via $E(B-V)$ and the extinction curve of \citet{Cardelli89}.  In
addition to Galactic reddening, the colors of M31 star clusters will
also be reddened by dust within M31.  Unfortunately it is not
straightforward to make such a correction, and therefore no additional
correction will be applied.  In later sections attention will be
focused on the near--IR colors of M31 clusters, where dust corrections
should be small.
 
The RBC contains a number of quality flags.  Clusters are selected to
be definite globular clusters (classification flag =1), with a
definite confirmation (confirmation flag =1), and to not be in the
young cluster catalog of \citet{FusiPecci05}.  This sample is further
restricted to clusters with $E(B-V)<0.2$ to reduce the uncertainties
associated with Galactic reddening.

The final cluster catalog, which includes the aforementioned quality
cuts and metallicity measurements, contains 22 clusters, 20 of which
include {\it GALEX NUV} photometry, and 12 of which include {\it GALEX
  FUV} photometry.

\subsubsection{The Magellanic Clouds}
\label{s:mcdat}

Data on star clusters in the MCs come principally from two sources.
\citet{Pessev06} presented a carefully constructed catalog of 2MASS
$JHK$ photometry for a sample of 75 star clusters.  In
\citet{Pessev08} this near-IR photometry is combined with $BV$
photometry from the literature.  These authors construct a `test
sample' of star clusters where high quality data are available to
estimate main sequence turn-off ages from CMDs and metallicities from
either the CMDs or individual stellar spectroscopy.  These
high--quality data are then binned by age to produce average colors
and metallicities as a function of cluster age.

In the youngest age bin a less direct method is used for determining
cluster ages in the \citet{Pessev08} sample.  The `$S$-parameter',
introduced by \citet{Elson85}, provides a simple empirical relation
between cluster age and a combination of integrated $U-B$ and $B-V$
colors.  High quality age estimates from HST observations from
\citet{Kerber07} were used by \citet{Pessev08} to update the
$S$-parameter--age calibration in order to provide age estimates for
the youngest clusters in their sample.  The Pessev et al. calibration
is substantially different from the commonly used relation of
\citet{Girardi95}.  It is important to keep in mind that since ages
are based on CMD fitting, they will depend on the stellar evolution
calculations used to convert main sequence turn--off points into ages.
In particular, the treatment of convective overshooting has a strong
impact on the age-- main sequence turn-off point relation for young
and intermediate ages ($t\lesssim2\times10^9$ yrs).  This sensitivity
arises because overshooting increases the amount of fuel available for
core hydrogen burning.  We will return to these issues when these data
are compared to models.

The second significant source of data on MC star clusters comes from
\citet{Gonzalez04}.  These authors analyze 2MASS data to produce not
only integrated near-IR colors but also surface brightness
fluctuations in the 2MASS $JHK$ filters.  Clusters are stacked
according to the classification system devised by
\citet[][SWB]{Searle80}, which closely tracks the S-parameter
described above.  The relation between SWB type, age and metallicity
is adopted from \citet{Cohen82}.  We have independently verified that
these ages agree with the ages inferred from the $S$-parameter
calibration discussed in \citet{Pessev08}.  This sample has been
augmented with $V-I$ colors \citep{Gonzalez05b}, where $I$ magnitudes
are from the DENIS survey \citep{Epchtein97} and $V$ magnitudes are
taken from \citet{vandenBergh81}.  We point out that the Pessev et
al. and Gonzalez et al. samples are not disjoint.  Comparison of these
two catalogs will provide a valuable cross--check on the reliability
of the data.

The novel feature of these two datasets is that the cluster data have
been stacked in bins of age in order to reduce stochastic effects.  A
number of authors have discussed the importance of stochastic effects
arising from small numbers of very bright stars \citep{Lancon00,
  Bruzual02, Cervino04, Bruzual03}.  Both \citet{Lancon00} and
\citet{Cervino04} provide limits on the minimum cluster mass that
should be satisfied so that stochastic fluctuations in integrated
colors are minimized.  \citet{Pessev08} demonstrate that the masses of
their stacked clusters easily satisfy both constraints, ensuring that
their stacked results can be robustly compared to SPS models.  The
masses of the stacked clusters in the \citet{Gonzalez04} sample are
comparable to those in \citet{Pessev08}, and so their results should
also be relatively unaffected by stochastic effects.

Finally, we include for completeness the $V-K$ and $J-K$ colors of LMC
clusters from \citet{Persson83}, which have been used in many
comparisons between observations and SPS models.  These $J$ and $K$
magnitudes are converted from the California Institute of Technology
(CIT) photometric system to the 2MASS system using the transformations
provided by \citet{Carpenter01}.  Ages for these clusters were
assigned based on the $S$-parameter calibration of \citet{Pessev08}.

\citet{Pessev06} presented a detailed comparison between their near-IR
colors and those of \citet{Persson83}, in some cases noting
substantial differences in the sense that colors from Persson et
al. were redder than the colors from Pessev et al.  They attributed
this difference to the fact that Persson et al. centered their
photometry on the brightest part of the cluster in order to maximize
the measured flux, while Pessev et al. center their apertures using a
Gaussian smoothing technique.  In some cases the Persson et
al. centroids are significantly offset from the true dynamical center
of the cluster because of off--center IR--bright sources within the
cluster.  For this reason, in addition to the fact that the data are
stacked to reduce stochastic effects, the results of Pessev et
al. should be preferred to the older results of Persson et al.  We
will return to this issue in $\S$\ref{s:calib}.

\subsection{Galaxy data}
\label{s:gal}

\subsubsection{The utility of galaxies}

For the reasons mentioned in the previous section, star clusters are
the ideal observational data to calibrate SPS models.  There are
however at least two reasons why it is useful to also consider SPS
constraints from galaxies: 1) galaxies tend to be of higher
metallicity than old star clusters, and so SPS model predictions for
old $\sim Z_\Sol$ populations can be assessed with galaxies composed
of old stars; 2) galaxies contain many more stars than star clusters,
and so stochastic effects are never a source of concern when comparing
models to galaxy data.

These two benefits of using galaxies to calibrate SPS models are
exploited herein by considering two classes of galaxies.  The first
class consists of bright red sequence, `quiescent' galaxies, which
contain old metal--rich stars, and will therefore be useful for
calibrating old metal--rich SPS model predictions.  The second class
are called `K+A', or `post--starburst' galaxies, as their spectra are
dominated by a mixture of A and K type stars and show no signs of
ongoing star formation \citep[e.g.,][]{Dressler83, Goto03}.  Such
galaxies are presumed to have had their star formation abruptly
truncated.  The presence of A stars in their spectra indicates that
the galaxy contains a significant population of $\lesssim1$ Gyr old
stars \citep[e.g.,][]{Leonardi96}.  These galaxies will provide
valuable constraints on the importance of TP--AGB stars, as the
prominence of this stellar phase peaks for ages $\approx1-2$ Gyr
\citep{Maraston05, Conroy09a}.  Since stars spend a small amount of
time in this phase, observational constraints from star clusters can
be difficult to interpret because of stochastic variations.  As
mentioned above, a stacking technique is employed when comparing to
star clusters so that stochastic effects are minimized.  Nonetheless,
it will be useful to derive independent constraints from galaxies
where stochastic effects are known to be negligible.

As mentioned in the introduction, the use of galaxies to assess the
accuracy of SPS models is complicated by at least two effects: the
attenuation of starlight by dust and the fact that galaxies contain
stars of a range of ages and metallicities.  These complications are
addressed where relevant in $\S$\ref{s:calib}.

\subsubsection{Data sources}

The source of all galaxy positions, optical photometry, and redshifts
is the Sloan Digital Sky Survey \citep[SDSS;][]{York00}, as made
available through the hybrid NYU Value Added Galaxy
Catalog\footnote{\texttt{http://sdss.physics.nyu.edu/vagc/}}
\citep[VAGC;][]{Blanton05}.  The SDSS photometry is k--corrected and
corrected for Galactic extinction via the \texttt{kcorrect v4.1.4}
software package \citep{Blanton07}.  The photometry is k--corrected to
$z=0.1$ -- the median redshift of the sample -- in order to minimize
the magnitude of the k--correction.  Spectroscopic indices are
provided by the MPA/JHU
database\footnote{\texttt{http://www.mpa-garching.mpg.de/SDSS/}}.

Near--IR photometry in $YJHK$ filters is provided by the UKIRT
Infrared Deep Sky Survey\footnote{\texttt{http://www.ukidss.org/}}
(UKIDSS).  The UKIDSS project is defined in \citet{Lawrence07}. UKIDSS
uses the UKIRT Wide Field Camera \citep[WFCAM;][]{Casali07}. The
photometric system is described in \citet{Hewett06}, and the
calibration is described in \citet{Hodgkin09}. The pipeline processing
and science archive are described in Irwin et al (2009, in prep) and
\citet{Hambly08}. We make use of Data Release 3, which is publically
available.  UKIDSS photometry is matched to the SDSS with a search
radius of 5''.  Petrosian magnitudes are used for the UKIDSS
photometry.  We do not consider colors that are combinations of SDSS
and UKIDSS photometry, so it is not essential that the photometric
apertures be precisely matched.

The UKIDSS survey goes much deeper than 2MASS, and is therefore more
desirable.  For example, typical 2MASS photometric errors for galaxies
at $z\sim0.1$ are 0.2 mags, compared to $\sim0.01$ for UKIDSS
photometry.  Galactic extinction is corrected using $E(B-V)$ reddening
estimates \citep{Schlegel98} in conjunction with a \citet{Cardelli89}
extinction curve.  K--corrections for the UKIDSS photometry are
estimated from the \texttt{kcorrect v4.1.4} software package using the
2MASS $JHK$ filters since those, and not UKIDSS filters, are available
in the code.  The filter transmission curves for 2MASS and UKIDSS are
very similar and so this approximation is justified.

Three galaxy samples are considered.  The first are K+A galaxies as
identified by \citet{Quintero04}.  This sample contains 1194 galaxies.
Of these, only 186 have UKIDSS photometry.  When considering near--IR
colors of the K+A sample we use only these 186 galaxies, while when
SDSS colors are considered we use the full sample.

The second sample consists of bright red sequence (`quiescent')
galaxies.  This sample is defined by $M_r^{0.1}<-20.5$, where the
superscript indicates that the magnitude has been k--corrected to
$z=0.1$, and
\begin{equation}
(g-r)^{0.1}> -0.03\,(M_r^{0.1}+23.0)+0.94,
\end{equation}
which divides galaxies by the valley visible in the color--magnitude
diagram and effectively isolates red sequence galaxies.  Applying
these cuts and requiring that galaxies have both SDSS and UKIDSS
photometry results in a sample of 5340 galaxies at $0.08<z<0.12$.

The third sample is derived from the spectroscopic indices catalog.
Galaxies are selected with $0.05<z<0.1$ and with a median
spectroscopic signal--to--noise of $S/N>25$.  From this catalog two
classes of galaxies are defined: those with velocity dispersions
measured from the stellar continuum of $100<\sigma<200\kms$ and those
with $\sigma>200\kms$.  These samples contain 13214 and 6774 galaxies,
respectively.

%-----------------------------------------------------------------

\section{SPS Model Construction}
\label{s:model}

\subsection{Flexible SPS}
\label{s:fsps}

\subsubsection{Theoretical isochrones}

SPS model construction closely follows that of \citet{Conroy09a}, to
which the reader is referred for details.  In brief, this SPS code
(referred to below as FSPS, for `flexible' SPS) combines stellar
evolution calculations with stellar spectral libraries to produce
simple stellar populations (SSPs), which describe the evolution in
time of the spectrum of a mono--metallic, coeval set of stars.  The
initial stellar mass function (IMF) specifies the weights given to
stars of various masses.  The IMF of \citet{Kroupa01} is adopted
throughout.  FSPS also contains a number of phenomenological dust
attenuation prescriptions and produces spectra for composite stellar
populations.  Our SPS code is open--source and publicly
available\footnote{\texttt{www.astro.princeton.edu/$\sim$cconroy/SPS/}}.

The novel feature of this SPS model is that the essential aspects of
SPS are integrated in a flexible way so that the user may choose, for
example, their preferred set of isochrones and stellar spectral
libraries.  In addition, the user may alter the weights given to and
physical parameters of various advanced stages of stellar evolution
(see below), and may specify the IMF of their choosing.  The details
of these various model choices are now described.

We make use of both the
Padova\footnote{\texttt{http://stev.oapd.inaf.it/cgi-bin/cmd}}
\citep{Girardi00, Marigo07a, Marigo08} and
BaSTI\footnote{\texttt{http://www.oa-teramo.inaf.it/BASTI}}
\citep{Pietrinferni04, Cordier07} stellar evolution calculations, both
in the versions available as of November 2009.  These two groups
follow stellar evolution from the main sequence through the
thermally--pulsating asymptotic giant branch (TP--AGB) phase, although
the treatment of the TP--AGB phase is very different between the two
groups.

\begin{figure}[!t]
\begin{center}
\resizebox{3.7in}{!}{\includegraphics{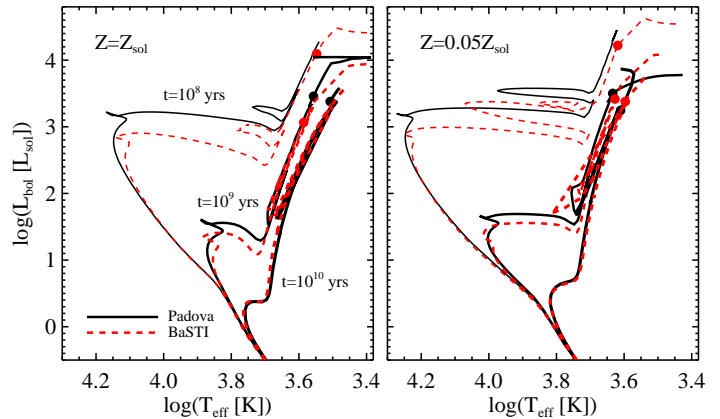}}
\end{center}
\vspace{0.5cm}
\caption{Theoretical HR diagram comparing the BaSTI and Padova
  evolutionary calculations at several ages and metallicities.
  Isochrones are plotted from the main sequence through the TP--AGB
  phase.  The onset of the TP--AGB phase is marked by solid symbols
  along each isochrone.  The $10^{10}$ yr solar metallicity BaSTI
  isochrone and the $10^8$ yr Padova isochrones do not contain TP--AGB
  stars.  The difference between the models at young ages is due to
  the different treatments of convection in the two models (i.e., the
  handling of convective core overshooting).}
\label{fig:cmd}
\vspace{0.5cm}
\end{figure}

The Padova calculations exist for metallicities in the range
$2\times10^{-4}<Z<0.030$, for ages $10^{6.6}<t<10^{10.2} $ yrs, and
for initial masses $0.15\leq M\leq67 \,\Msun$.  BaSTI outputs cover a
somewhat narrow range in ages, $10^{7.5}<t<10^{10.2} $ yrs, and
initial masses $0.5\leq M\leq10.0 \,\Msun$ but include higher
metallicities: $3\times10^{-4}<Z<0.040$.  The BaSTI database provides
models for both solar--scaled and $\alpha-$enhanced chemical
compositions, for two assumptions regarding convective core
overshooting, and for two stellar mass-loss rates.  We use the
solar--scaled, no overshooting, and $\eta=0.4$ mass--loss rate
parameter model.  For both lower and higher masses the Padova
calculations are stitched onto the BaSTI grid.  The Padova and BaSTI
models are compared in Figure \ref{fig:cmd}.  In this figure
isochrones are compared at several ages and metallicities, and in
addition the location of the TP--AGB onset is shown.

In order to follow stellar evolution from the end of the TP--AGB phase
through the post--AGB phase, the above evolutionary tracks have been
supplemented with the post--AGB models of \citet{Vassiliadis94}.  The
contribution of post--AGB stars to the integrated SED is highly
uncertain for a number of reasons, including a lack of calibrating
observations and the unknown duration of heavy enshroudment as
post--AGB stars leave their M giant phase.  Recent observations of M32
have found that post--AGB stars are much less frequent than standard
models predict \citep{Brown08b}.  Nonetheless, the inclusion of
post--AGB tracks may be important for the interpretation of the
ultraviolet spectra of old systems \citep[e.g.,][]{Greggio90}.

The FSPS code contains a number of parameters that control the weights
given to various uncertain phases of stellar evolution.  For example,
the bolometric luminosity and effective temperature of TP--AGB stars
can be arbitrarily modified.  These modifications are applied as
overall shifts in log($L_{\rm bol}$) and log($T_{\rm eff}$) for the
entire TP--AGB phase, and are parameterized by the variables
$\Delta_L$ and $\Delta_T$ \citep[see discussion in][]{Conroy09a}.  In
addition, an extended blue HB can be included\footnote{When we
  consider blue/extended HB populations we do not {\it add} new stars
  but rather move stars from the red clump HB to the blue component.},
as can a population of BS stars.  A blue HB population is
parameterized as the fraction, $f_{\rm BHB}$ of HB stars that are in a
blue/extended component.  Such a flexible treatment is motivated by
the great variety of observed HB morphologies in MW clusters, even at
a given metallicity \citep{Piotto02}.  These stars are uniformly
populated from the red clump to log$(T_{\rm eff}/K)=4.2$.  BS stars
are uniformly populated from one half magnitude above the main
sequence turn--off to two and a half magnitudes brighter than the
turn--off, along the zero--age main sequence \citep{Xin05}.  The
weight given to BS stars is specified as the number of BS stars per
unit HB star, $S_{\rm BS}$.  The weight given to the post--AGB phase
can also be arbitrarily modified.  The effect of these uncertain
aspects of stellar evolution will be investigated in later sections.

\subsubsection{Spectral libraries}

Stellar spectral libraries are required to convert the outputs of
stellar evolution calculations --- surface gravities, $g$, effective
temperatures, $T_{\rm eff}$, and metallicities, $Z$, --- into
observable spectral energy distributions.  Two spectral libraries are
considered herein.  The first is the empirically--calibrated
theoretical BaSeL3.1 library \citep{Lejeune97, Lejeune98, Westera02},
supplemented with empirical, average TP--AGB spectra from the library
of \citet{Lancon02}.  The BaSeL library was constructed from the model
atmospheres of Kurucz (1995, private communication to R. Buser),
\citet{Fluks94}, \citet{Bessell89b, Bessell91}, and \citet{Allard95}.
In the TP--AGB library, spectra are binned both by near--IR color and
by whether the star is carbon--rich or oxygen--rich.  Stars hotter
than $50,000K$ are assumed to be pure blackbodies.  This combined
library covers the full wavelength range (ultraviolet through
infrared) at a sampling of 10\AA\, in the ultraviolet and 20\AA\, in
the optical.

The second spectral library considered herein is derived from the
empirical Miles library \citep{Sanchez-Blazquez06}.  This atlas
contains spectra of $\approx1000$ stars covering the wavelength range
$3600<\lambda<7500$\AA\, at a sampling of 0.9\AA\ per pixel
(FWHM$\approx2.3$\AA).  Derived stellar parameters are presented in
\citet{Cenarro07}.  Integrating empirical libraries into SPS models is
a non--trivial task because the coverage in the stellar parameters
$g$, $T_{\rm eff}$, and $Z$ is very inhomogeneous.  In fact many
regions of this parameter space are not covered by empirical libraries
at all, especially at sub--solar and super--solar metallicities and
for spectral types earlier than A.  An additional problem with the
Miles library is that the spectra have been arbitrarily normalized to
unity at 5350\AA.  This is particularly problematic because the
wavelength range is not broad enough to compare the integral of the
flux to the bolometric luminosity for a star with a given $g$, $T_{\rm
  eff}$, and $Z$.  Therefore, an additional library must be used to
provide the absolute flux scale.  The BaSeL library is used for this
purpose.  The BaSeL library is also used to provide the relative
trends of spectra with stellar parameters for regions of parameter
space where the relative trends cannot be self--consistently
determined from Miles.  The Miles library is binned and, when
necessary, linearly interpolated onto the Kurucz/BaSeL
log$(g)$--log$(T_{\rm eff})$ grid.  This interpolated library is then
interfaced with FSPS in a manner identical to the BaSeL library.  Five
metallicity bins are defined so that roughly similar numbers of stars
occupy each bin.  The bin boundaries are:
log$(Z/Z_\Sol)=[-2.0,-1.0,-0.5,-0.15,0.15,0.75]$.  The resulting
average metallicities within these bins are
log$(Z/Z_\Sol)=[-1.4,-0.7,-0.3,0.0,0.2]$.

At metallicities less than solar, the Miles library stars are
$\alpha-$enhanced \citep{Milone09}.  At [Fe/H]$<-1$ the library is on
average enhanced in Mg by +0.4 dex.  This will be relevant in our
comparison to the spectra of MW star clusters in later sections.

There are several important caveats to these libraries that the reader
should keep in mind.  In the empirical TP--AGB library, the
metallicities of the stars are {\it unknown}.  The authors advocate
assuming that they are all solar metallicity since a significant
fraction are located in the solar neighborhood (although many are also
located in the MCs).  In order to use the library at non--solar
metallicities, the authors suggest the use of the theoretical
metallicity--dependent color--temperature relations in
\citet{Bessell91} in order to convert the observed $I-K$ colors into
temperatures at other metallicities.

\citet{Westera02} found that the original theoretical atmosphere
calculations used in the BaSeL library could not simultaneously match
observed $UBVRIJHKL$ color--temperature relations for individual stars
and CMDs of globular clusters.  The reason for this was thought to be
inadequacies in the stellar evolution calculations used in the CMD
comparisons.  Because of this problem, \citet{Westera02} presented two
normalizations of the BaSeL library, one which matched the
color--temperature relations, and the other which matched the globular
cluster data.  These two normalizations are referred to as the `WLBC'
and PADOVA2000' libraries, since the Padova evolutionary calculations
were used in the globular cluster comparisons.  The `WLBC'
normalization will be used herein because this calibration is not tied
to any particular isochrone set.

In summary, we have in hand a flexible SPS model from which we can
choose two sets of isochrones (Padova or BaSTI) and two sets of
spectral libraries (BaSeL or Miles).  In addition, there are a number
of free parameters characterizing various uncertain aspects of
advanced stages of stellar evolution.  These various options will be
exploited in later sections.

\subsubsection{Modifications to the Padova isochrones}
\label{s:mod}

For the Padova isochrones, the TP--AGB phase has been modified in
order to generate acceptable agreement with the data.  Specifically,
the TP--AGB phase has been calibrated against both MC star clusters
and post--starburst galaxies ($\S$\ref{s:cmc} and $\S$\ref{s:cpsg},
respectively).  These calibrations result in an adjustment of the
TP--AGB parameters $\Delta_L$ and $\Delta_T$ (quantifying overall
shifts in log$(L_{\rm bol})$ and log$(T_{\rm eff})$; see
$\S$\ref{s:fsps}) as follows:
\begin{equation}
\Delta_L  =\left\{ \begin{array}{ll} - 1.0+(t-8.0)/1.5, & 8<t<9.1
    \\[1mm] -{\rm max}\{{\rm min}\{0.4,-{\rm log}(Z/Z_\Sol)\},0.2\}, & t>9.1,
\end{array}
\right.
\end{equation}
\begin{equation}
  \Delta_T  =\left\{ \begin{array}{ll} \left\{ \begin{array}{ll}
          +0.1,& {\rm log}(Z/Z_\Sol)<-0.25 \\ 0.0, & {\rm otherwise} \end{array}\right. & 8<t<9.1
      \\[1mm]  +0.1-{\rm min}\{(t-9.1)/1.5,0.2\}, & t>9.1,
\end{array}
\right.
\end{equation}
where $t$ is the age of the population measured in log(yrs).  We
emphasize that the values of $\Delta_L$ and $\Delta_T$ chosen to fit
the data are {\it not unique}, and that, unless explicitly stated in
the above definitions, they are independent of
metallicity. Throughout, reference to the `modified Padova model'
refers to these modifications.

\subsection{Additional SPS models}

\begin{figure*}[!t]
\begin{center}
\resizebox{5in}{!}{\includegraphics{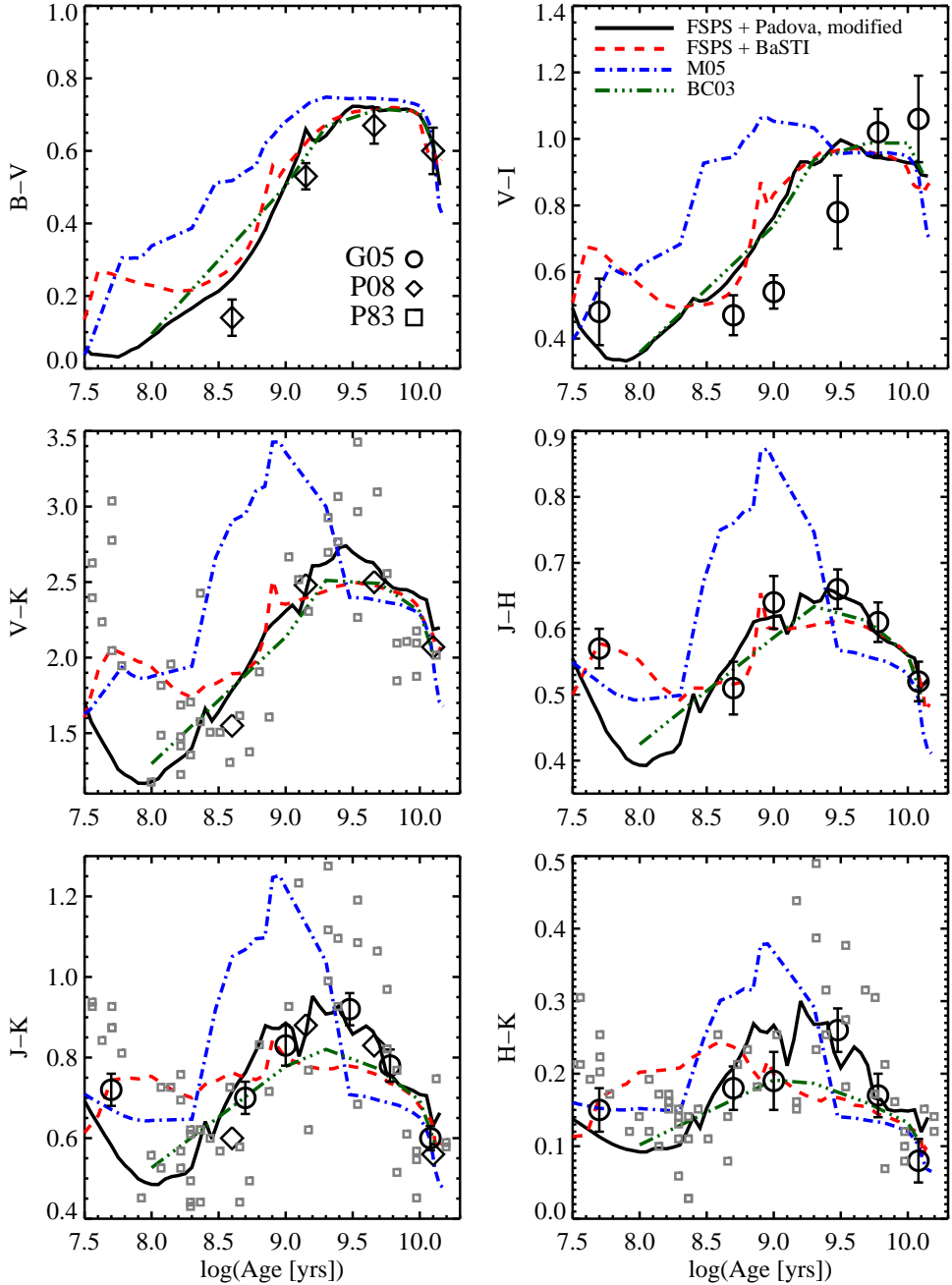}}
\end{center}
\vspace{1.cm}
\caption{SSP colors as a function of age, comparing a variety of SPS
  models to observations.  Data are derived from MC star clusters.
  The model SSPs include the predictions from M05 ({\it dot--dashed
    lines}), BC03 ({\it dot-dot-dot-dashed lines}) and two models
  derived from the present work.  Using our FSPS code, a model was
  constructed using the BaSTI ({\it dashed lines}) and the modified
  Padova stellar evolution calculations ({\it solid lines}).  The MC
  data are from \citet[][P83]{Persson83}, \citet[][G05]{Gonzalez05b}
  and \citet[][P08]{Pessev08}.  For the latter two data sets, each
  data point represents an average over many observed star clusters
  within each age bin.  The Persson data is believed to be less
  reliable for reasons described in the text.  The average metallicity
  of the data decreases as the age increases, from
  log$(Z/Z_\Sol)=-0.3$ at log$(t)$=7.5 years to log$(Z/Z_\Sol)=-1.5$
  at log$(t)$=10.0 years.  The model predictions reflect this change
  in average metallicity with age.}
\label{fig:mccol1}
\vspace{0.5cm}
\end{figure*}

We now describe the salient aspects of the BC03 and M05 SPS models.

The BC03 model utilizes the Padova isochrones circa 1994
\citep{Alongi93, Bressan93, Fagotto94}, supplemented with the
\citet{Vassiliadis93} models for TP--AGB stars and the
\citet{Vassiliadis94} models for post--AGB stars.  BC03 use the models of
\citet{Groenewegen93} to differentiate between carbon--rich and
oxygen--rich TP-AGB stars.  There are no blue/extended HB stars in the
BC03 models.  Their models are constructed with both the theoretical
BaSeL library (described in the previous section) and the empirical
STELIB library \citep{LeBorgne03}.  Spectra for carbon stars are taken
from model atmosphere calculations \citep{Hofner00}.

The empirical STELIB library contains optical spectra for 249 stars at
a resolving power of $R\approx2000$ (FWHM$\sim3$\AA).  The effective
temperatures in the STELIB library are not reliable and so BC03 adopt
the color--temperature scale from the BaSeL library in order to relate
the stellar evolutionary outputs to the empirical library.  There are
very few stars at non--solar metallicities and so the reliability of
models constructed with this library at non--solar metallicities is
questionable.  Indeed, \citet{Koleva08} compared several different SPS
models and found that the BC03 model constructed with STELIB produces
unreliable results at non--solar metallicities.

We make use of the BC03 model that was constructed with the STELIB
library for the optical spectra.  For both ultraviolet and near--IR
predictions, the STELIB library has been extended with the BaSeL
library.

The M05 model is built from the following ingredients.  Stellar
evolution calculations through the main sequence turnoff are adopted
from Cassisi et al. \citep{Cassisi97a, Cassisi00}.  The fuel
consumption theorem \citep{Renzini86, Maraston98} is then used to
incorporate advanced stages of stellar evolution through the TP--AGB
phase.  The evolutionary calculations are converted to observables
with the theoretical BaSeL spectral library for all stars except
TP--AGB stars.  For these stars the empirical library of
\citet{Lancon02} is used (this is the same library used in our FSPS
code).

Model magnitudes and spectral indices\footnote{Spectroscopic indices
  are defined by the wavelength intervals specified in the MPA/JHU
  database.  The intervals are similar to the classic Lick/IDS system
  \citep{Worthey94b}, although no attempt has been made to ensure that
our indices are on the Lick system scale.}
are computed self--consistently from the model spectra.  Comparisons
between models in later sections are therefore not plagued by
potentially different filter transmission curves or index definitions.

\begin{figure}[!t]
\begin{center}
\resizebox{3.2in}{!}{\includegraphics{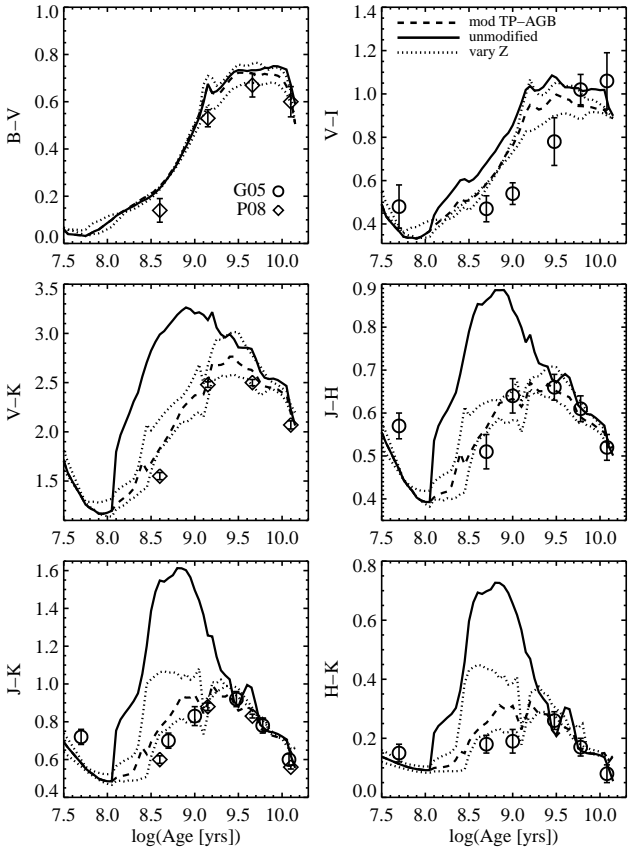}}
\end{center}
\vspace{.5cm}
\caption{SSP colors as a function of age, demonstrating the
  sensitivity to the TP--AGB phase.  The Padova stellar evolution
  calculations are used for the model predictions.  The modified model
  shown in Figure \ref{fig:mccol1} ({\it dashed lines}) is compared to
  the unmodified Padova calculations ({\it solid lines}).  Dotted
  lines correspond to a $\pm0.2$ dex variation in metallicity about
  the modified model.  The data are from star clusters in the MCs, as
  in Figure \ref{fig:mccol1}.  Notice the different $y-$axis ranges in
  the bottom panels compared to Figure \ref{fig:mccol1}.}
\label{fig:mccol2}
\vspace{0.5cm}
\end{figure}

%-----------------------------------------------------------------

\section{SPS Model Calibration \& Comparison}
\label{s:calib}

\subsection{Overview}

We now turn to a comparison between observations and SPS models.  In
$\S$\ref{s:cmc}--\ref{s:cmwm31} we focus on star clusters, both in the
MCs, where constraints are strong for intermediate age, sub--solar
metallicities, and in the MW and M31, where constraints are strong for
old metal--poor systems.  We then turn to galaxies in
$\S$\ref{s:cqg}--\ref{s:cpsg}.  Two types of galaxies will be
considered: massive red sequence galaxies and post--starburst
galaxies.  The former class will provide a qualitative assessment of
the accuracy the SPS models in the solar to super--solar metallicity,
old age regime.  The post--starburst galaxies will provide an
assessment of the models in the approximately solar metallicity,
intermediate age regime.  As discussed in previous sections, it is
challenging to use galaxies to assess the accuracy of SPS models
because galaxies are composed of stars with a range of ages and
metallicities and dust reddening can be important.  These
complications are discussed extensively in
$\S$\ref{s:cqg}--\ref{s:cpsg}.

\begin{figure*}[!t]
\begin{center}
\resizebox{4.5in}{!}{\includegraphics{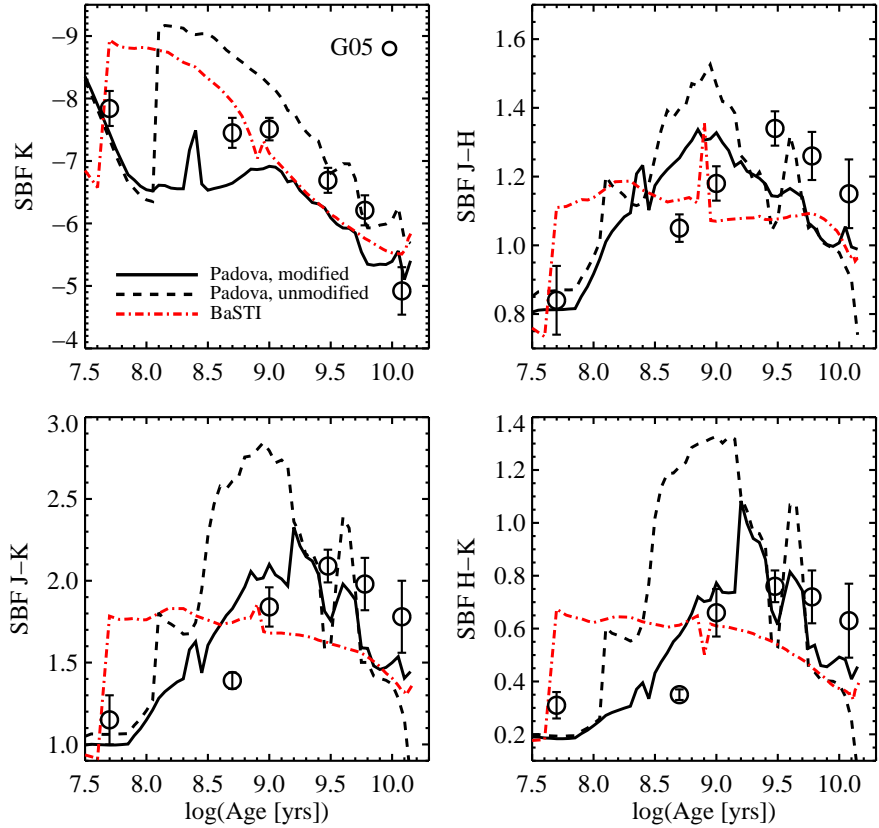}}
\end{center}
\vspace{0.2cm}
\caption{SSP surface brightness fluctuation (SBF) magnitudes and
  colors as a function of age.  Models are compared to data from star
  clusters in the MCs \citep[][G05; {\it symbols}]{Gonzalez05b}.  The
  models are based on our FSPS code using the Padova and BaSTI
  isochrones.  For the Padova isochrones, predictions are included
  both for the unmodified and modified isochrones.}
\label{fig:sbf1}
\vspace{0.2cm}
\end{figure*}

\subsection{Constraints from the Magellanic Clouds}
\label{s:cmc}

\subsubsection{Colors}
\label{s:mccol}

Figure \ref{fig:mccol1} shows the colors of MC star clusters as a
function of cluster age.  As discussed in $\S$\ref{s:mcdat},
individual clusters are binned in age to produce `superclusters' that
are relatively immune to stochastic fluctuations \citep{Pessev08,
  Gonzalez05b}.  The data from \citet{Persson83} are also included,
and have not been binned in age.  These data are of inferior quality
(see $\S$\ref{s:mcdat}) and are only included for comparison because
this data is frequently used to constrain SPS models
\citep[e.g.,][]{Maraston05}.

A variety of SPS models are compared to the MC star cluster data in
Figure \ref{fig:mccol1}.  From our FSPS code we have generated SSP
color evolution for both the Padova and BaSTI isochrones.  In this
figure we make use of the modified Padova isochrones (see
$\S$\ref{s:mod}).  The models of BC03 and M05 are also included.  All
model predictions reflect the fact that the average metallicity of the
data varies with age.  That is, we have interpolated model predictions
so that they vary with metallicity in as in the data, where the
age--metallicity relation in the MC data is
log($t$/yrs)$=(7.0,7.7,8.7,9.0,9.5,9.8,10.1)$,
log$(Z/Z_\Sol)=(-0.28,-0.28,-0.28,-0.28,-0.38,-0.68,-1.5)$
\citep{Pessev08, Gonzalez05b}.

\begin{figure}[!t]
\begin{center}
\resizebox{2.5in}{!}{\includegraphics{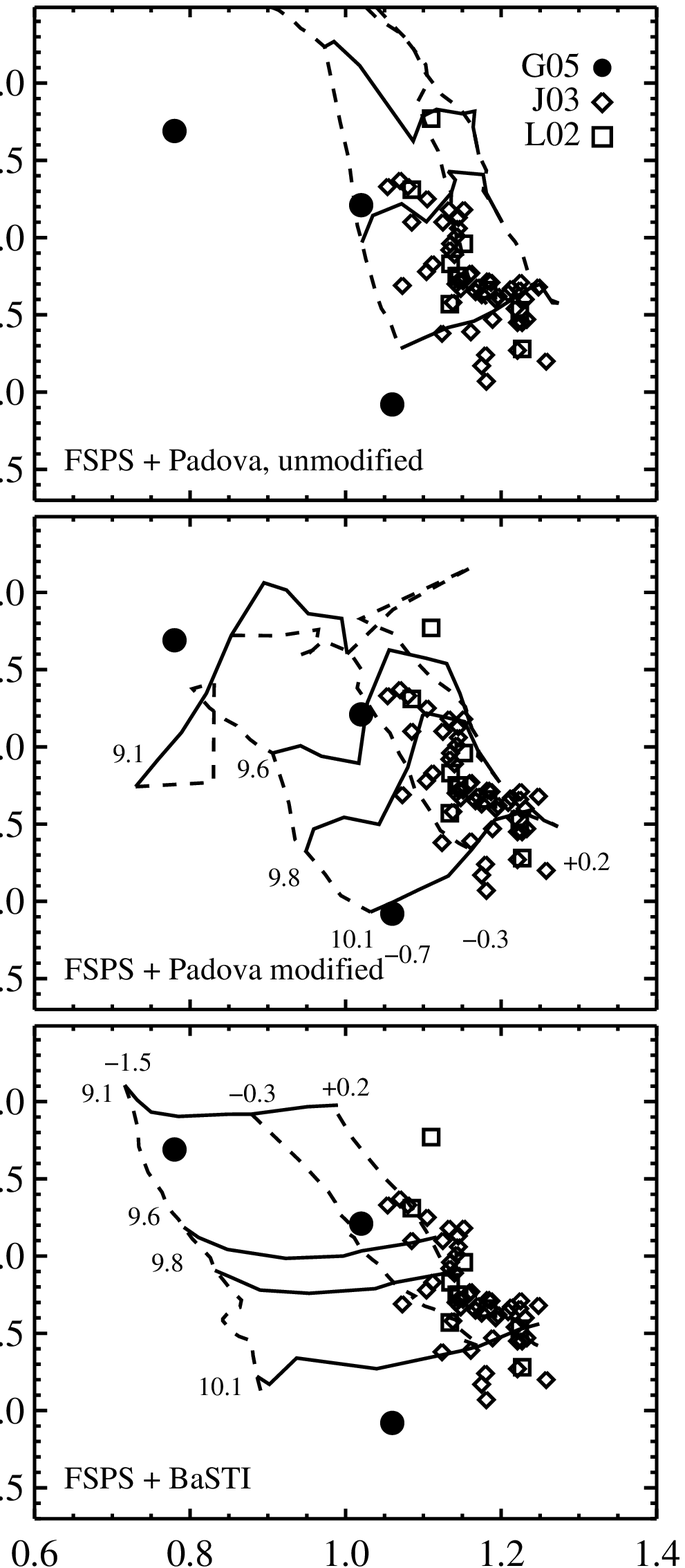}}
\end{center}
\vspace{0.2cm}
\caption{$K-$band SBFs as a function of integrated $V-I$ color.  Model
  predictions ({\it lines}) are compared to observations of star
  clusters in the MCs \citep[][G05]{Gonzalez05b}, Elliptical galaxies
  in the Fornax cluster \citep[][L02]{Liu02}, and Ellipticals in a
  variety of environments \citep[][J03]{Jensen03}.  Model predictions
  are displayed as lines of constant age ({\it solid lines}; labeled
  in units of log$(t/{\rm yrs})$ in the figure) and lines of constant
  metallicity ({\it dashed lines)}; labeled in units of
  log$(Z/Z_\Sol)$ in the figure).  The age--metallicity grids in the
  top two panels are the same.  SSP models are constructed with the
  unmodified Padova isochrones ({\it top panel}), modified Padova
  isochrones ({\it middle panel}), and the BaSTI isochrones ({\it
    bottom panel}).}
\label{fig:sbf2}
\vspace{0.4cm}
\end{figure}

The M05 models are significantly redder than the data at intermediate
ages, owing to the different TP--AGB treatment in M05 \citep[see
also][who reach similar conclusions]{Pessev08}.  The M05 colors also
have an age--dependence that does not agree with the data.  The MC
cluster ages were estimated via CMD fitting using isochrones that
include convective overshooting \citep{Kerber07, Pessev08}, whereas
the M05 models do not include overshooting, and so one might expect
some mismatch in age between the M05 models and data.  However, the
differences induced by the treatment of overshooting on the MC ages is
$\lesssim0.2$ dex at ages $10^8\lesssim t\lesssim10^9$ yrs, whereas
the discrepancy in the age--dependence between M05 and the data is
$>0.5$ dex in age.  We therefore regard most of this discrepancy as a
shortcoming of the M05 model, with only a fraction being due to
different definitions of cluster age.

The Persson et al. data are substantially redder than the newer
datasets from Pessev et al. and Gonzalez--Lopezlira et al.  As
discussed in \citet{Pessev06}, this is probably due to the fact that
the Persson photometry was probably centered on the brightest star in
the cluster, which are often carbon stars, and in any case are very
red, and so the photometry will be biased red.  The difference between
Persson's data and the 2MASS--based data highlights the care that must
be taken to construct datasets that can be meaningfully used for SPS
model calibration.

The M05 models were calibrated against color--color plots of MC star
clusters using the Persson et al. data and very wide age bins.
Stochastic effects will tend to smear star clusters along an underling
color--color relation because if a cluster is scattered redder in one
color, it will be redder in another color as well.  The M05 model
calibration is therefore sensitive to stochastic effects, and is in
addition plagued by the systematics in the Persson data.  Moreover,
the wide age bins used in M05 explains why the discrepancies in age
seen for the M05 model in Figure \ref{fig:mccol1} were not seen in
their own calibrations.

The BC03 model and BaSTI isochrones both fare well except for $H-K$
and $J-K$ colors at intermediate ages (log$(t/{\rm yrs})\approx9.5$),
for which models are too blue by $\approx0.1$ mags compared to the
data.  This disagreement is probably due to the simplistic treatment
of TP--AGB stars in these models.  The BaSTI isochrones also produce
redder colors at young ages compared to the Padova isochrones, which
is probably due a lack of convective core overshooting in the BaSTI
isochrones adopted herein (see Figure \ref{fig:cmd}).  The data do not
definitively favor one model over another at log$(t/{\rm
  yrs})\lesssim8.0$.

The disagreement in $V-I$ between the models and data at intermediate
ages is also noteworthy.  The models perform better at old ages, although
it is unclear if this is because of the old ages or the lower
metallicities of the old star clusters.  We will return to issues with
$V-I$ colors in later sections.

FSPS model predictions using the Padova isochrones are explored
further in Figure \ref{fig:mccol2}.  In this figure we compare the
unmodified Padova isochrones to the modified model, and we also show
the effect of a variation in metallicity of $\pm0.2$ dex about the
modified model.  For ages and colors this variation with metallicity
does not bracket the modified model because some metallicity--color
relations in the models are not monotonic and in fact some colors are
reddest at approximately the average metallicity of the LMC
(log$(Z/Z_\Sol)\approx-0.3$).  It is clear that use of the unmodified
Padova models results in substantial disagreement with observed star
clusters.  This disagreement is driven to a significant extent by the
large population of luminous carbon stars predicted in the Padova
calculations.  In our modifications to the Padova models we have
substantially lowered the bolometric luminosity along the TP--AGB
phase, which has the effect of down--weighting the importance of all
TP--AGB stars, including carbon stars.

\begin{figure*}[!t]
\begin{center}
\resizebox{5.5in}{!}{\includegraphics{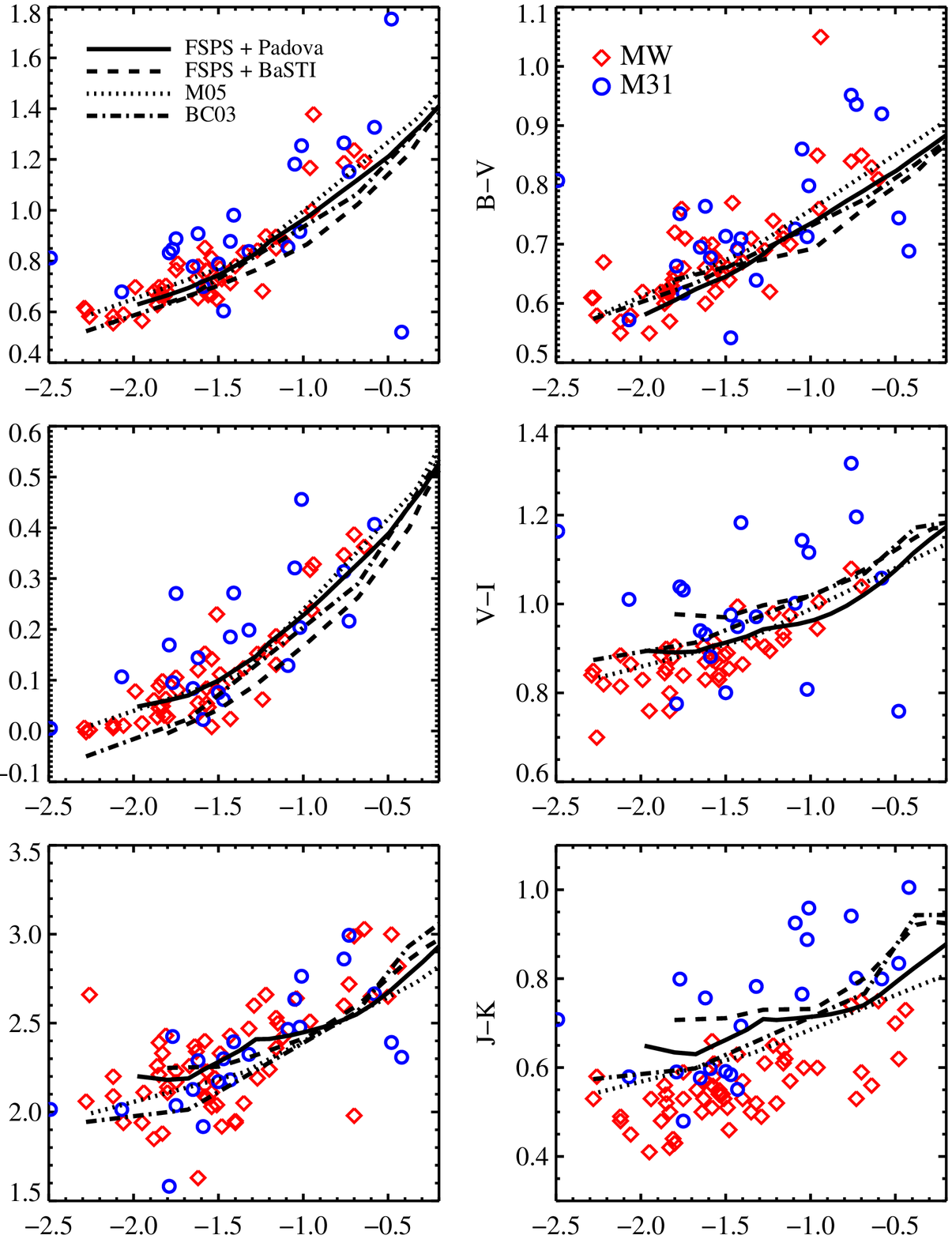}}
\end{center}
\vspace{0.5cm}
\caption{SSP colors as a function of metallicity, for old ages
  ($t\approx13$ Gyr).  Model predictions are compared to observations
  of globular clusters in the MW and M31.  The model SSPs include the
  predictions from M05 ({\it dotted lines}), BC03 ({\it dot-dashed
    lines}), and predictions from the present work where either the
  Padova or BaSTI stellar evolution calculations have been assumed.
  In general the agreement between models and data is acceptable,
  though the data tend to be redder than the models as the metallicity
  increases.  The difference in $J-K$ color between MW and M31 star
  clusters is striking.}
\label{fig:lZ}
\vspace{0.5cm}
\end{figure*}

\begin{figure*}[!t]
\begin{center}
\resizebox{5.5in}{!}{\includegraphics{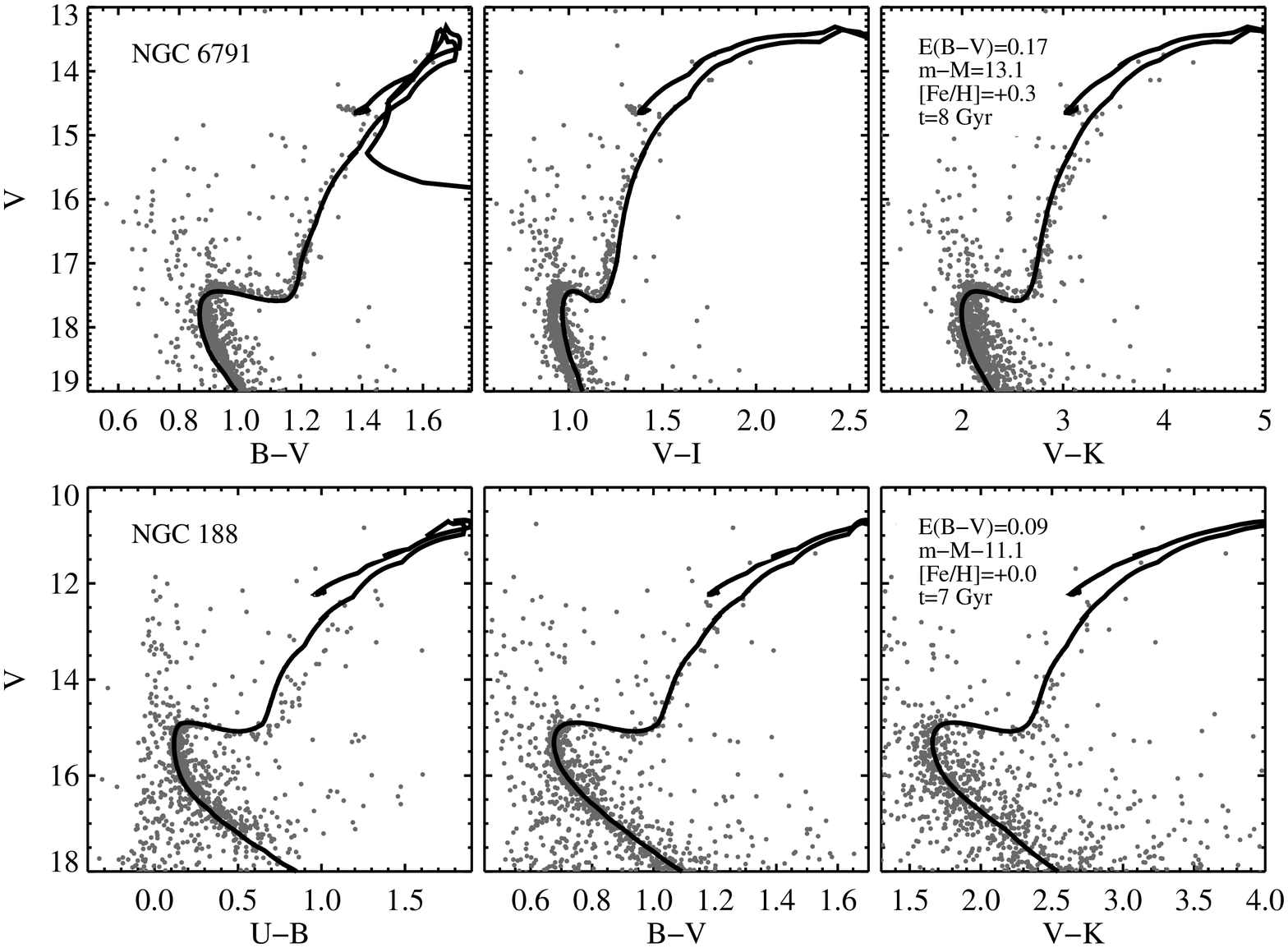}}
\end{center}
\vspace{.5cm}
\caption{CMDs for two metal--rich MW star clusters, NGC 6791 ({\it top
    rows}) and NGC 188 ({\it bottom rows}), compared to FSPS model
  predictions using the BaSTI isochrones.  Notice the strong BS
  population above the main sequence turn--off for both NGC 6791 and
  NGC 188.  The adopted cluster parameters are shown in the legends.}
\label{fig:hZ}
\vspace{0.2cm}
\end{figure*}

It is worth discussing in more detail the substantial disagreement
between the unmodified Padova calculations and the star cluster data.
The Padova group provided comparisons between their model and
observations in \citet{Marigo08} for $V-K$ and $J-K$ colors, finding
acceptable agreement.  The explanation for this difference is twofold.

First, the data used in their comparison is the old Persson data,
which suffers from the issues discussed above.  Second, and perhaps as
important, the conversion from $L_{\rm bol}$ and $T_{\rm eff}$ to SEDs
was achieved with the theoretical atmosphere models of \citet{Loidl01}
coupled to dust radiative transfer models from \citet{Groenewegen06}.
Aside from the fact that the dust models were never demonstrated to
fit the $V-K$ colors of AGB stars, the \citet{Loidl01} carbon star
models are substantially bluer (by $>1$mag in $V-K$) than observed
carbon stars (R. Gautschy, private communication).  We believe that
our transformation to SEDs is more robust since we use observed
TP--AGB spectra.  Our qualitative conclusions are echoed by
\citet{Gullieuszik08} who compared the population of O--rich and
carbon stars in the Leo II dwarf galaxy to the Padova models and found
that the models predict substantially more, and substantially more
luminous O--rich stars than are seen in that galaxy.  Finally, the
Padova group uses the carbon star luminosity function in the MCs as a
constraint on their TP--AGB model, but this observable can only
provide valuable insight if the star formation histories (SFHs) of the
MCs are known to high precision in the age range relevant for carbon
star production ($\approx10^{8.5}-10^{9.15}$ yrs).

\subsubsection{Surface brightness fluctuations}

Figures \ref{fig:sbf1} and \ref{fig:sbf2} compare near--IR SBFs
between several FSPS--based models and MC data.  In Figure
\ref{fig:sbf1} the evolution of SBF magnitudes and SBF colors are
compared to our FSPS model with both the BaSTI and Padova isochrones.
We show results both for the unmodified and modified Padova
isochrones.  The BaSTI isochrones provide an acceptable fit to the
$K-$band SBFs, but produce SBF colors that are somewhat too blue
compared to the data.  This implies that the AGB and TP--AGB
temperatures are too hot in the BaSTI calculations.

The unmodified Padova results show similar discrepancies as seen in
Figure \ref{fig:mccol2}.  SBF magnitudes and colors are too bright and
too red, which can be traced back to the over--abundance of carbon
stars and the over--luminous TP--AGB evolutionary phase.  As discussed
earlier in $\S$\ref{s:mccol}, these problems are rectified by altering the
overall luminosity and temperature of stars in the TP--AGB phase.
These modifications result in SBF magnitudes and colors that are in
better accord with the data, although $K-$band SBFs tend to be
somewhat lower than the observations at intermediate ages.

\begin{figure}[!t]
\begin{center}
\resizebox{2.5in}{!}{\includegraphics{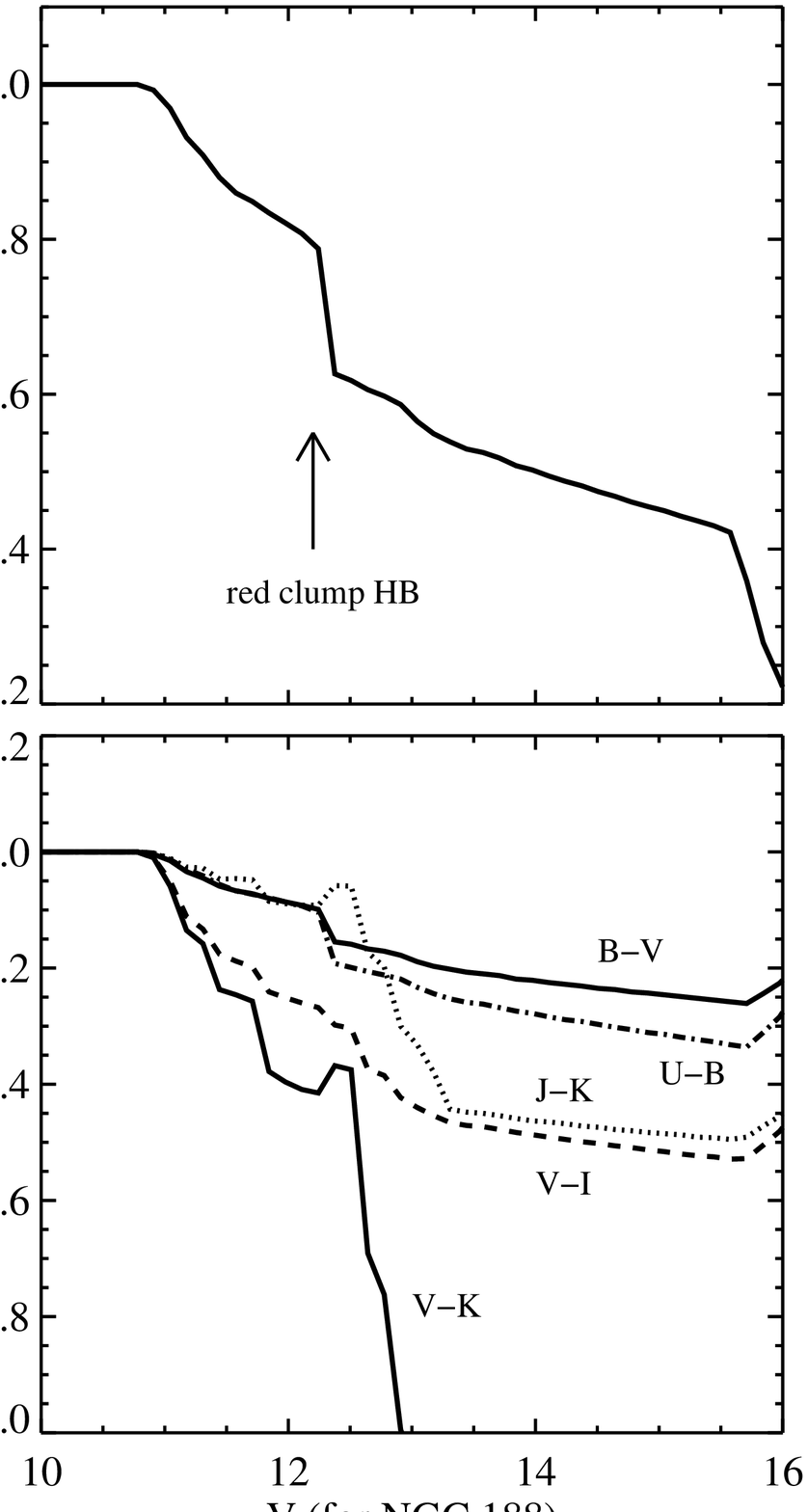}}
\end{center}
\vspace{.5cm}
\caption{{\it Top panel:} Cumulative fraction of $V-$band light as a
  function of apparent $V-$band magnitude for a 7 Gyr SSP with
  [Fe/H]$=+0.0$.  The apparent $V-$band magnitude along the $x-$axis
  is scaled by the distance modulus and reddening value appropriate
  for the star cluster NGC 188 shown in Figure \ref{fig:hZ}.  Red
  clump HB stars appear at $V\sim12$.  {\it Bottom panel:} Integrated
  color for all stars fainter than $V$, normalized to the integrated
  color for all stars.  The model SSP is the same as in the top panel.
  Integrated colors in the red and near--IR can only be constrained
  from CMDs if the very brightest and rarest stars in the cluster are
  identified.}
\label{fig:CMDf}
\vspace{0.5cm}
\end{figure}

In Figure \ref{fig:sbf2} we compare models and data in $K-$band SBFs
versus integrated $V-I$ colors.  In addition to MC star cluster data
we include data from Elliptical galaxies \citep{Liu02, Jensen03}.  It
is clear that the modified Padova isochrones span the range of data
much better than the unmodified isochrones.  In addition, notice that
there is a subsample of Ellipticals that have SBFs fainter than
predicted by the models.  This has a direct and potentially important
impact on the inferred mass--to--light ratios of these systems and
consequently on their inferred stellar masses.  SBFs in the
red/near--IR will correlate inversely with derived stellar masses, and
therefore if SBFs are overestimated in the models (as is the case
here), then derived stellar masses will be underestimated.

The models shown in Figure \ref{fig:sbf2} assume solar--scaled
chemical compositions, while many Elliptical galaxies are
$\alpha-$enhanced \citep[e.g.,][]{Worthey92, Thomas05}.  However,
\citet{LeeHC09c} have shown that the level of $\alpha-$enhancement has
almost no effect on the model predictions in the $V-I$ vs. $F160W$ SBF
plane ($F160W$ is similar to the $H-$band), and therefore our
conclusions should not be effected by these details.

\begin{figure}[!t]
\begin{center}
\resizebox{3.3in}{!}{\includegraphics{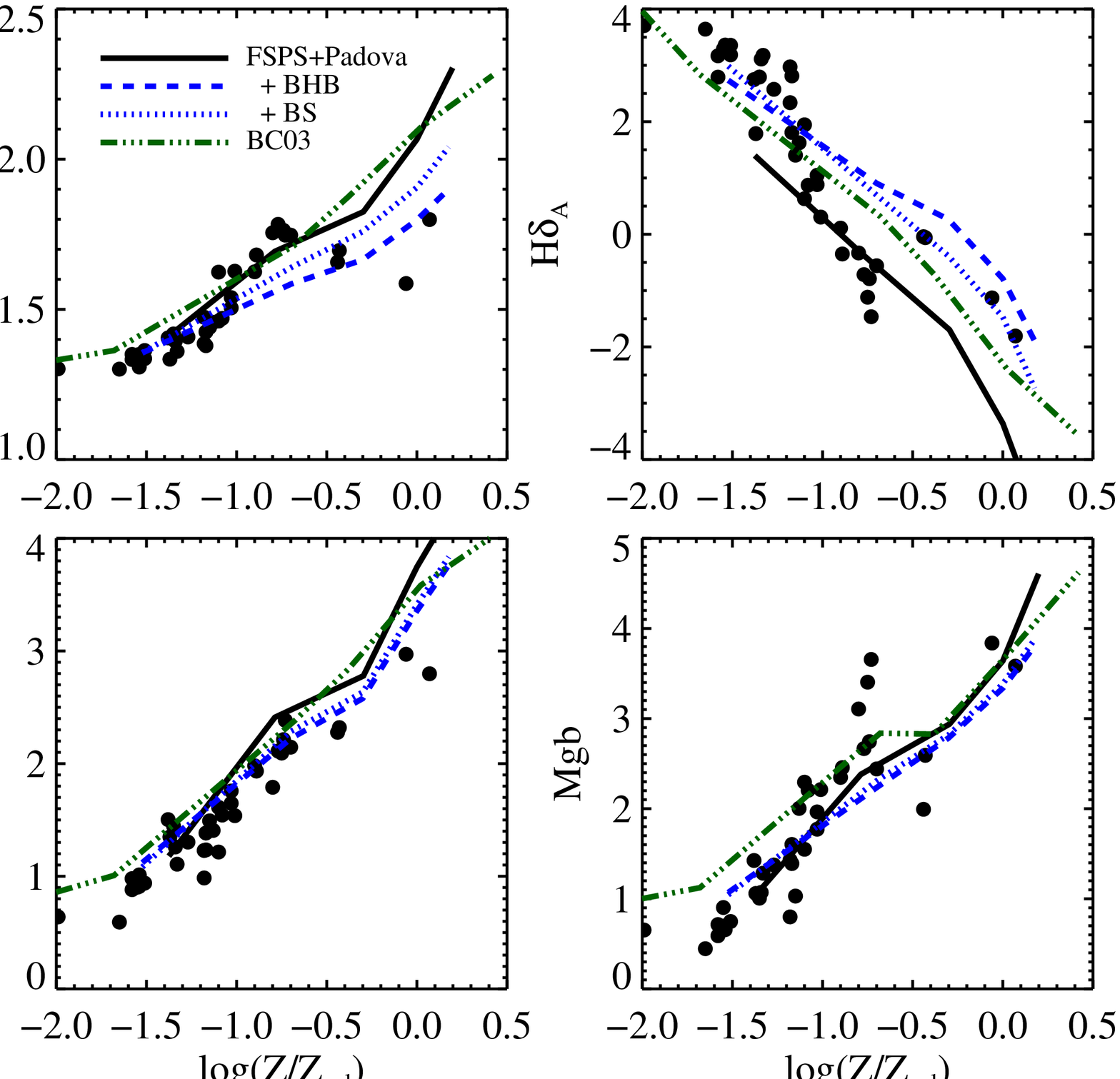}}
\end{center}
\vspace{.2cm}
\caption{Spectral indices versus metallicity for old populations.
  Indices are derived from the integrated spectra of MW star clusters
  presented in \citet[][{\it symbols}]{Schiavon05}, and are compared
  to indices of SSPs at $t=10$ Gyr for several SPS models.  Here, our
  FSPS model uses Padova isochrones and the empirical Miles stellar
  library.  Additional FSPS predictions that include a fraction of
  blue HB stars ($f_{\rm BHB}=0.5$) and BS stars ($S_{\rm BS}=2$) are
  shown.  The BaSTI isochrones (not shown), are similar in most
  respects to the Padova isochrones, except for the metal--poor
  H$\delta_A$ strengths, which are larger in the BaSTI models because
  of a bluer HB.  The model predictions of BC03 are also included in
  the figure.}
\label{fig:MWlick}
\vspace{0.5cm}
\end{figure}

\begin{figure*}[!t]
\begin{center}
\resizebox{5in}{!}{\includegraphics{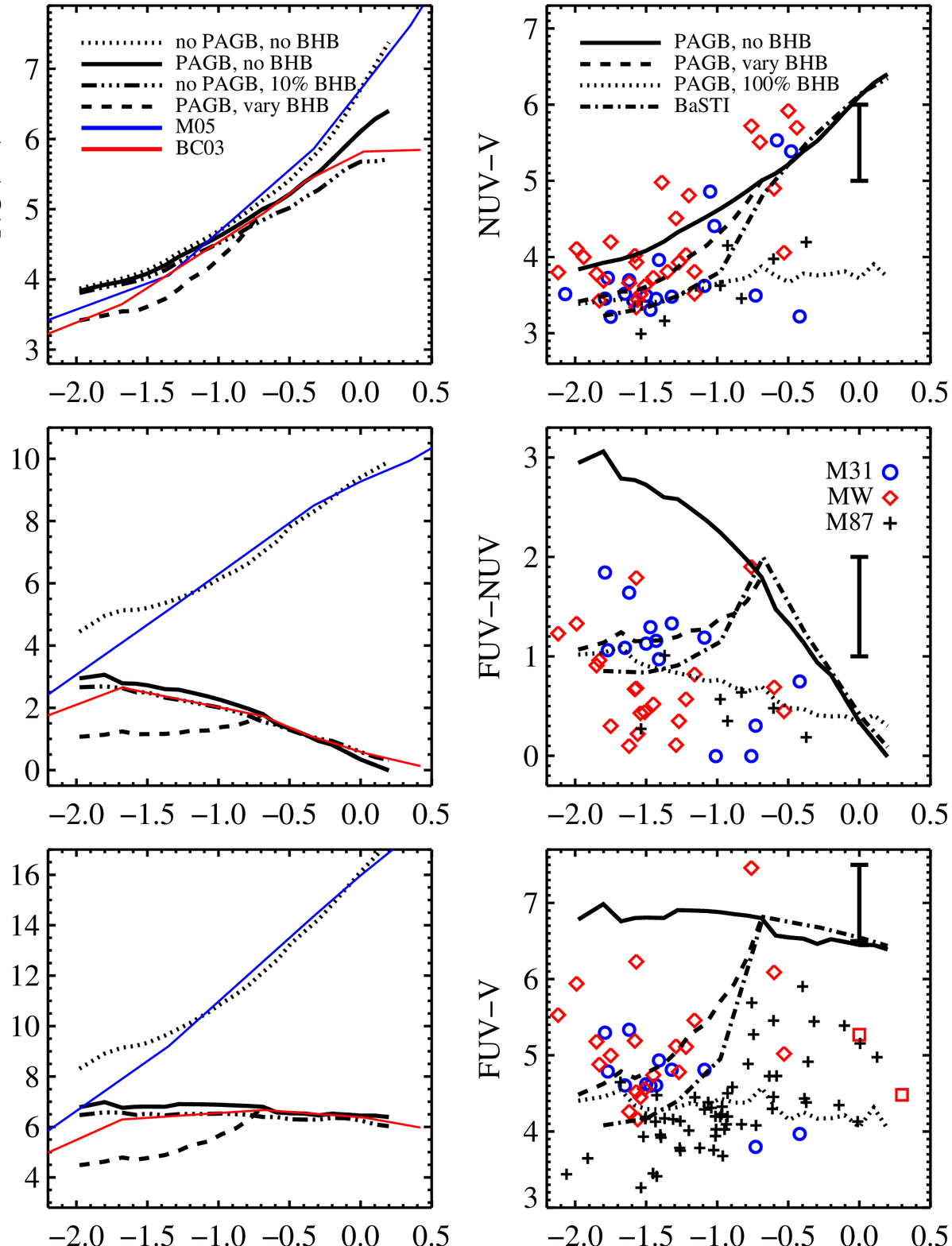}}
\end{center}
\vspace{.5cm}
\caption{SSP UV colors as a function of metallicity, for old
  populations ($t\approx13$ Gyr).  Model predictions are compared to
  observations of star clusters in M31 \citep{Rey07}, the MW
  \citep{Dorman95}, and M87 \citep{Sohn06}.  A variety of models are
  considered to explore the effects of post--AGB (PAGB) stars and blue
  HB (BHB) stars.  These models, which are described in detail in the
  text, are constructed with the Padova isochrones.  Model predictions
  from M05, BC03 are also included, as is a model with the BaSTI
  isochrones.  In addition, the approximate location of Elliptical
  galaxies from the SAURON survey \citep{Jeong09} are included in the
  right set of panels as error bars at log$(Z/Z_\Sun)=0.0$.
  Approximate locations of two metal--rich MW star clusters NGC 188
  and NGC 6791 are shown as open squares \citep[see][]{Sohn06}.
  Notice the different $y-$axes in the left and right panels.}
\label{fig:uv}
\vspace{0.4cm}
\end{figure*}

\subsection{Constraints from the MW \& M31}
\label{s:cmwm31}

\subsubsection{Optical and near--IR colors}
\label{s:mwopt}

SPS model predictions for the metallicity--dependent optical and
near--IR colors of old populations are compared to observed star
clusters in Figure \ref{fig:lZ}.  Model predictions are restricted to
ages of 12.5 Gyr, although results are not sensitive to this
particular age.  FSPS models with both the BaSTI and modified Padova
isochrones are included, as are the predictions from M05 and BC03.
Comparisons to star clusters in both the MW and M31 are included.  For
all colors except for $J-K$ the MW and M31 data agree (see below),
although the scatter in M31 cluster colors is larger.  This is
probably due to the effects of internal reddening in M31 that has not
been accounted for.

The models reproduce the trends in the observed colors farely well,
although there are areas of noteworthy disagreement.  The $U-V$ and
$B-V$ colors are somewhat too blue in the models at
log$(Z/Z_\Sol)>-1.0$.  The FSPS model with the BaSTI isochrones
perform worst of the models considered.  We can check whether or not
this general disagreement between models and data in $UBV$ filters is
due to the adopted spectral library by considering the empirical Miles
stellar spectral library.  Due to the limited wavelength range of this
library, we can only construct the $B-V$ color.  The $B-V$ colors from
the Miles empirical library are almost indistinguishable from the
colors using the BaSeL library, implying that the stellar atmospheres
are not the source of the discrepancy.  A similar test was performed
in BC03, where they compare the BaSeL, STELIB, and \citet{Pickles98}
libraries and find negligible differences in $B-V$ at $Z=Z_\Sol$.

A second possible explanation for the differences between models and
observations in the $U-V$ and $B-V$ colors is the $UBV$ filter
transmission curves.  The $B$ filter used for these data actually
consist of two slightly different filters \citep[see the discussion
in][]{Maraston05}.  It is difficult to determine which filter was used
in each observation.  The difference between these two filters varies
from 0.02 to 0.04 mags.  The $U-V$ color is constructed by adding the
$U-B$ and $B-V$ colors, so this color will also be subject to this
source of uncertainty.  The offsets between models and data is of
order 0.05 mags --- the same order as the differences induced by the
two $B$ filters --- and so we cannot make stronger comparisons between
models and data until more reliable optical photometry is available
for MW and M31 star clusters.

The morphology of the HB cannot explain the observed discrepancy
because in all SPS models shown in Figure \ref{fig:lZ} adopt a red HB
at log$(Z/Z_\Sol)>-1.0$.  The HB therefore cannot be modified to
produce even redder model colors.

The model $V-I$ colors compare favorably with the data at
log$(Z/Z_\Sol)>-1.0$.  At lower metallicities the models diverge from
the data, with the FSPS model using BaSTI isochrones performing worst.

All models fail at predicting the observed $V-K$ and $J-K$ colors,
both at low and high metallicities for the latter and at higher
metallicities for the former.  Recall that the near--IR data comes
from a recent analysis of 2MASS data by \citet{Cohen07}.  Our results
echo those of \citet{Cohen07} who compared MW star cluster $J-K$
colors against several SPS models.  In addition to the M05 model, they
also compare to the models of \citet{Worthey94} and \citet{Buzzoni89}
and find similar levels of disagreement with these other models.

\citet{Cohen07} speculate that the cause of the discrepancy could be
either 1) inadequate calibration of the stellar libraries or 2) issues
with the near--IR filter transmission curves (Cohen et al. used
published predictions for Johnson $JHK$ photometry from these models).
Since we compute 2MASS magnitudes self--consistently from the model
spectra, the second explanation cannot be the source of the
discrepancy.  The problem is not likely to be due to inadequacies in
the theoretical stellar spectral libraries, since the colors of giants
compare well with observed stars in the near--IR \citep{Westera02}.
The temperature of the RGB need only be lowered by 200K in order to
achieve better agreement in $V-K$ (although a lower RGB temperature
would only exacerbate the disagreement in $J-K$).  As explored in
$\S$\ref{s:metalr}, one need only change the RGB at the brightest
$\sim1$ mag to in order to produce substantial changes in $V-K$
colors.  Without definitively identifying the source of this problem,
we are forced to simply reiterate the conclusions of Cohen et al. that
near--IR colors of old populations cannot be used to constrain the
metallicities or ages of either GCs or distant galaxies.

It should also be noted that the near--IR photometry presented in
Cohen et al., which is based on data from the 2MASS survey, differs
substantially from previous near--IR photometry of MW clusters
provided by M. Aaronson et al. (1977; unpublished).  The differences
in $J-K$ are $0.1-0.2$ mags in the sense that the latest results are
bluer than the older data.  Since previous models were calibrated
against the older data \citep[e.g.,][]{Maraston05}, it is not
surprising that the models are all redder than the Cohen et al. data.

The most dramatic discrepancy in Figure \ref{fig:lZ} is between the
$J-K$ colors in MW and M31 star clusters.  The M31 clusters are
approximately 0.2 mags redder in $J-K$ than the MW clusters.  Internal
reddening in M31 cannot be the source of this discrepancy because
reddening effects are quite small in $J-K$, especially in comparison
to the colors considered in Figure \ref{fig:lZ}, which do not show
such an offset.  \citet{Barmby00} have also presented $JHK$ photometry
of old M31 star clusters.  Their results are consistent with the M31
results from the RBC catalog shown in Figure \ref{fig:lZ}, suggesting
that the discrepancy in $J-K$ is not due to data reduction issues.  We
cannot provide a compelling explanation for this discrepancy, but note
that without an explanation it is impossible to calibrate the old
metal--poor models to better than $\sim0.2$ mags in the near--IR.

\vspace{0.4cm}
\subsubsection{Metal--rich CMDs}
\label{s:metalr}

Calibrating data at higher metallicities than shown in Figure
\ref{fig:lZ} is far more desirable from the standpoint of using SPS to
model galaxies since galaxies are generally of higher metallicity.
Unfortunately, old metal--rich star clusters are rare.  Classic
metal--rich clusters include NGC 6791, NGC 188, NGC 6553, and NGC
6528; the latter two belong to the bulge and thus are along sightlines
of high extinction.

In Figure \ref{fig:hZ} we compare multi--color CMDs of NGC 6791 and
NGC 188 to our SPS predictions.  The consideration of multi--color
CMDs provides a powerful constraint on SPS models in part because the
reddening value cannot be arbitrarily varied to fit a color without
potentially worsening the fit at another color.  NGC 188 and 6791 are
open clusters in crowded fields and therefore CMD construction is
challenging without radial velocities and/or proper motions to isolate
true members.  This complication should be kept in mind when
considering model comparisons below.

NGC 6791 is assumed to have an age of 8.5 Gyr, $E(B-V)=0.17$,
[Fe/H]$=+0.4$, [$\alpha$/Fe]$=0.0$, and $m-M=13.0$ \citep{Carney05,
  Origlia06, Kalirai07}.  $JK$ and $BVI$ photometry are provided by
\citet{Carney05} and \citet{Stetson03}, respectively.  The cluster
parameters for NGC 188 are taken to be $E(B-V)=0.09$, [Fe/H]$=-0.06$,
[$\alpha$/Fe]$=0.0$, $m-M=11.1$, and an age of $7$ Gyr
\citep{Sarajedini99, Carretta09}.  $UBVRI$ photometry of this cluster
is provided by \citet{Sarajedini99}; near--IR photometry is extracted
directly from the 2MASS point source catalog\footnote{The 2MASS point
  source catalog was queried using the NASA/IPAC Infrared Science
  Archive, available at
  \texttt{http://irsa.ipac.caltech.edu/applications/Gator/.}}.

We use the BaSTI isochrones for the following comparisons because
those isochrones have been computed for log$(Z/Z_\Sol)=+0.3$, which is
comparable to the metallicity of NGC 6791 and is higher than the
highest metallicities available from the Padova isochrones.  Our model
CMDs are compared to NGC 6791 and NGC 188 in Figure \ref{fig:hZ} for
the same reddening values, ages, and distance moduli as quoted above,
but for slightly difference metallicities.  Metallicities of
log$(Z/Z_\Sol)=+0.3$ for NGC 6791 and log$(Z/Z_\Sol)=+0.0$ for NGC 188
are adopted based on the available isochrones.

The agreement between data and models is generally good, including the
location of the main sequence turn--off, the sub--giant branch and
RGB, and the position of the red HB.  

The model fails to reproduce the low mass end of the main sequence in
NGC 6791 in the sense that the model isochrone is on the blue side of
the data in $B-V$ and $V-K$ and on the red side in $V-I$.  Of course,
this disagreement is of little consequence for integrated properties
since low mass stars are faint.  The disagreement at low masses is
probably due to uncertainties associated with color--$T_{\rm eff}$
relations for low--mass stars at high metallicity \citep{Cassisi00,
  Cassisi07}. It appears that the model RGB $V-K$ color becomes
progressively bluer than the RGB of NGC 6791 at higher luminosities,
although it is difficult to make firm statements owing to the paucity
of data at the bright end.  Notice that this discrepancy is in the
same sense as was seen for integrated $V-K$ colors of metal--poor MW
and M31 clusters (see $\S$\ref{s:mwopt}).

The model CMDs perform well when compared to NGC 188.  The only obvious
discrepancy is in the $U-B$ color of the RGB.  Thankfully, this is of
little consequence for integrated colors because the integrated $U-B$
color is determined primarily by the main sequence turn--off point and
sub--giant branch, not the RGB (see below).

For the purpose of modeling entire galaxies, we are interested in the
ability of SPS models to accurately predict the {\it integrated}
colors of star clusters.  It is therefore worth considering the extent
to which constraints from resolved CMDs can constrain integrated
colors of clusters.  

In Figure \ref{fig:CMDf} the integrated colors corresponding to the
CMD of NGC 188 are explored.  Integrated colors are computed by
including all stars fainter than a given apparent $V-$band magnitude.
It is clear from this figure that integrated blue colors such as $U-B$
and $B-V$ are well constrained even at relatively faint magnitudes in
$V$.  This occurs because integrated blue colors are determined
primarily by the main sequence turn--off point and sub--giant branch.
In contrast, redder and near--IR colors are highly sensitive to the
brightest one magnitude ($11<V<12$) for the simple reason that the RGB
and AGB dominate the near--IR light.  The brightest one magnitude in
$V$ contributes $\approx20-40$\% of the integrated $V-$band light
(depending on whether or not one includes the red clump HB), and the
$V-I$ and $V-K$ colors redden dramatically in this brightest
magnitude.  Thus, unless the brightest, and rarest stars in clusters
can be reliably identified and used for model calibration, model
constraints in CMD--space will not guarantee accurate model integrated
colors.

\subsubsection{Spectral indices}

We now turn to an analysis of the spectral indices of old populations
over a range of metallicities.  Observed indices are measured from the
catalog of optical spectra for 40 MW star clusters from
\citet{Schiavon05}.  The indices are measured in the exact same way
for both the observed spectra and the models, ensuring reliable
comparison.  The FSPS model spectra are at slightly higher resolution
than the MW cluster spectra, so the model spectra have been broadened
to match the resolution of the data.  The spectral resolution of the
BC03 model is comparable to the observed spectra, and so the BC03
model spectra were not broadened.

Figure \ref{fig:MWlick} presents the spectral indices D$_n4000$,
H$\delta_A$, Fe5270, and Mgb as a function of metallicity for the
Schiavon et al. sample and the FSPS and BC03 SPS models.  The FSPS
predictions utilize the Miles empirical stellar library, while BC03
adopt the STELIB empirical library.  For these ages the Padova
predictions are very similar to the BaSTI isochrones except for
H$\delta_A$ at low metallicity because the BaSTI isochrones have a
hotter HB.  This figure also demonstrates the importance of BS stars
and an extended/blue HB by including models with $S_{\rm BS}=2$ and
$f_{\rm BHB}=0.5$ (see $\S$\ref{s:fsps}).

There are several important conclusions to draw from this figure.
First, the models are able to reproduce the observed metallicity
dependence of the Mgb and Fe5270 indices over the full observed range.
At low metallicities this success is driven largely by the fact that
the bulk of the metal--poor stars in the empirical stellar library
contains $\alpha-$enhanced stars \citep{Milone09}.  While the
isochrones used in FSPS are solar--scaled\footnote{Although recall
  that the BaSTI database does provide $\alpha-$enhanced stellar
  evolution models.}, recent work has shown that for old populations
the effects of $\alpha-$enhancement are strong on the stellar
libraries and weak on the isochrones \citep{Schiavon07, LeeHC09,
  LeeHC09b}.

The default SPS models (i.e., without BS or blue HB stars) predict
D$_n4000$ in excess of observed values at solar metallicity, and the
same models predict H$\delta_A$ strengths too weak compared to the
observations at both very low and solar metallicity.  The disagreement
at matching H$\delta_A$ at very low metallicity is a consequence of
the well--known fact that most low metallicity clusters have a blue
HB.  Indeed, the models do a better job of matching the observed
H$\delta_A$ strengths at low metallicity for $f_{\rm BHB}>0.5$.

The failure of the models for both D$_n4000$ and H$\delta_A$ at solar
metallicity is more striking, and has potentially significant
consequences for understanding the spectra of metal--rich galaxies.
The two solar metallicity clusters are NGC 6553 and NGC 6528; both are
in the bulge.  The CMD of NGC 6553 shows an abundance of BS stars
\citep{Zoccali01}, and so the addition of such stars to the models in
Figure \ref{fig:MWlick} is justified.  Indeed, adding BS stars results
in much better agreement in H$\delta_A$ and moderately better
agreement for D$_n4000$.  The addition of blue HB stars also
alleviates much of the discrepancy between models and data.
\citet{LeeHC09b} compare their model to the same dataset and find that
for solar--scaled models the H$\delta_A$ strengths are too weak
compared to the observed $Z\approx Z_\Sol$ clusters, echoing our
results regarding our default SPS model.

Main sequence turn-off stars are the dominant stellar phase
controlling the strength of D$_n4000$.  The effective temperature of
turn-off stars would have to be increased by $\sim1000$K in order to
alleviate the discrepancy at solar metallicity.  Such an increase
would yield a CMD in gross conflict with observations.  Modifications
to the standard stellar phases thus cannot reconcile the data and
model D$_n4000$ strengths.  Either BS, blue HB stars or both are
therefore the most likely candidates for explaining the discrepancy.

\subsubsection{Ultraviolet colors}
\label{s:cuv}

\begin{figure*}[!t]
\begin{center}
\resizebox{4.8in}{!}{\includegraphics{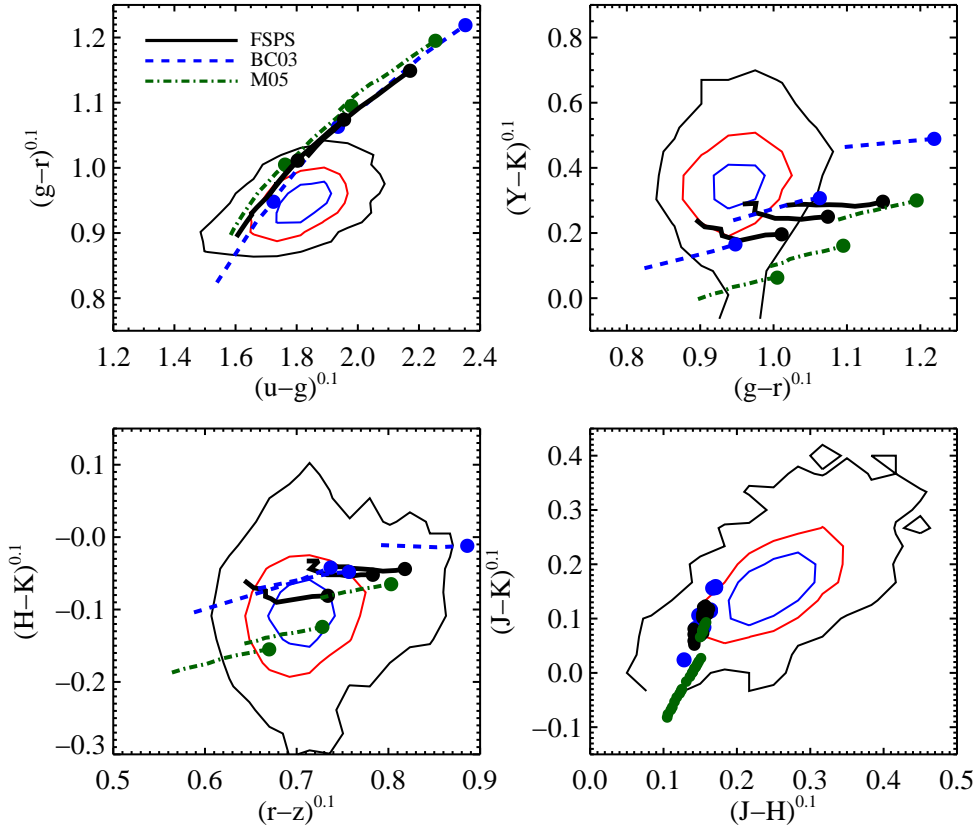}}
\end{center}
\vspace{.2cm}
\caption{Color--color diagrams comparing observed luminous red
  sequence galaxies to both our FSPS model with the modified Padova
  isochrones, and the models of BC03 and M05.  The contours enclose
  40\%, 68\%, and 95\% of the observed galaxies.  For each model, the
  colors of dust--free SSPs are plotted for ages $5<t<13$ Gyr for
  three metallicities: ${\rm log}(Z/Z_\Sol)\approx-0.2,0.0,0.2$
  (corresponding to three lines for each model).  Symbols mark the
  location of $t=13$ Gyr for each metallicity.  In the lower right
  panel, all model predictions are shown as symbols for clarity. }
\label{fig:redcol}
\vspace{0.4cm}
\end{figure*}

\begin{figure*}[!t]
\plotone{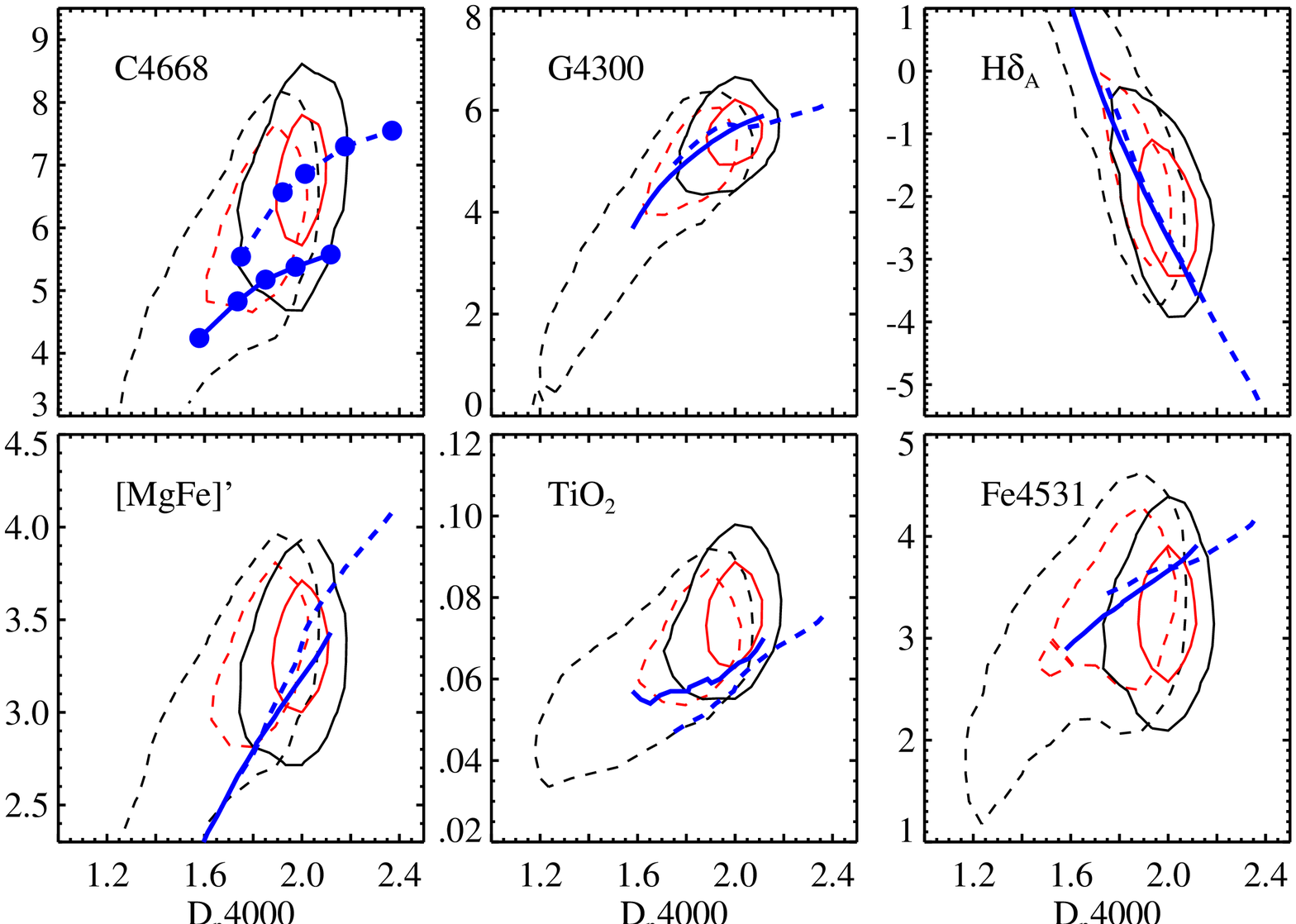}
\vspace{.8cm}
\caption{Comparison of spectral indices for a sample of SDSS galaxies
  ({\it 68\% and 95\% contours}) to our FSPS model predictions ({\it
    lines}).  Here, our FSPS model uses the empirical Miles stellar
  library and the Padova isochrones.  SDSS galaxies are divided by
  $\sigma>200\kms$ ({\it solid contours}) and $100<\sigma<200\kms$
  ({\it dashed contours}).  Model predictions are displayed for
  dust--free SSPs with metallicities of $Z_\Sol$ ({\it solid lines})
  and $1.5Z_\Sol$ ({\it dashed lines}) for $2<t<13$ Gyr.  Model
  spectra are Doppler broadened to $\sigma=200\kms$.  In the upper
  left panel, symbols denote ages of $2.2, 3.5, 5.6, 8.9, 14.1$ Gyr,
  in order of increasing $D_n4000$.}
\label{fig:lick1}
\vspace{0.5cm}
\end{figure*}

In the absence of evolved or young stars, the ultraviolet is sensitive
to the temperature of main sequence turn--off stars.  Since the
temperature of turn--off stars depends on metallicity at fixed age,
ultraviolet colors such as $NUV-V$ and $FUV-V$ will, in the absence of
evolved or young stars, redden substantially with increasing
metallicity.

In reality, the interpretation of ultraviolet colors of old
populations is complicated by the existence of hot, evolved stars,
including post--AGB, extended/blue HB, and BS stars
\citep[e.g.,][]{Burstein88, Greggio90, Dorman95, Oconnell99, Brown00,
  Han07}.  It is beyond the scope of this paper to consider in detail
the constraints on the relative importance of these various
populations afforded by ultraviolet colors.  Instead, our goal will be
to demonstrate that plausible variations in the relative weights
assigned to post--AGB and blue HB stars can explain the ultraviolet
properties of observed star clusters.

Ultraviolet colors of star clusters in M31 \citep{Rey07}, the MW
\citep[as compiled by][]{Dorman95}, and M87 \citep{Sohn06} are
compared to SPS models in Figure \ref{fig:uv}.  The approximate
location of Elliptical galaxies from the SAURON survey \citep{Jeong09}
are also included in the figure.  We include model predictions from
our FSPS code for both the Padova and BaSTI isochrones, along with
model predictions from BC03 and M05.  Recall that M05 does not include
post--AGB stars and so the FUV photometry, which is sensitive to such
hot stars, cannot be reliably compared to the data.  M05 does include
blue HB stars for metal--poor models.  BC03 includes post--AGB stars
but does not include blue HB stars at any metallicity.

For the FSPS predictions with the Padova isochrones, we include
several model variations.  In particular, we consider models 1)
without post--AGB and without blue HB stars; 2) with post--AGB but
without blue HB stars; 3) without post--AGB stars and with a blue HB
fraction of 10\%; 4) with post--AGB stars and a blue HB fraction of
100\%; 5) with post--AGB stars and a blue HB fraction that varies with
metallicity as $f_{\rm BHB}=-0.75{\rm log}(Z/Z_\Sol)-0.5$ for ${\rm
  log}(Z/Z_\Sol)<-0.7$ and zero otherwise.

\begin{figure*}[!t]
\plotone{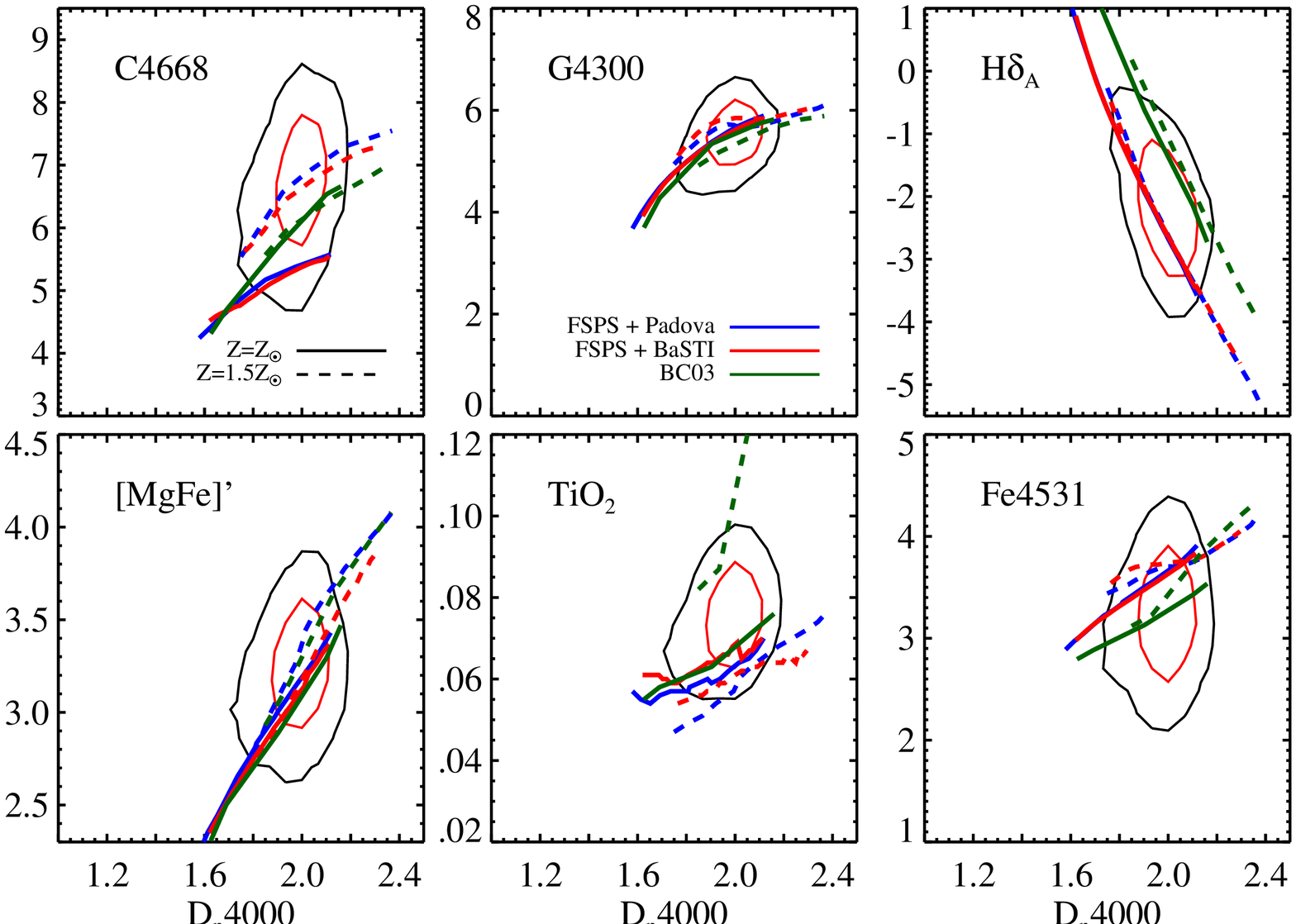}
\vspace{.8cm}
\caption{Spectral indices for a sample of SDSS galaxies with
  $\sigma>200\kms$ ({\it contours}), compared to several models for
  dust--free SSPs.  FSPS model predictions are shown for both the
  Padova and BaSTI isochrones where in each case the empirical Miles
  stellar library is used in the model construction.  Predictions from
  the BC03 model are also included.  Models are shown for metallicities
  of $Z_\Sol$ ({\it solid lines}) and $1.5Z_\Sol$ ({\it dashed lines})
  and for ages in the range $2<t<13$ Gyr.  Model spectra are
  Doppler broadened to $\sigma=200\kms$. }
\label{fig:lick2}
\vspace{0.5cm}
\end{figure*}

Before considering the insights gained from model comparisons, it is
worth drawing attention to the considerable variation in observed
ultraviolet colors of star clusters at fixed metallicity.  $NUV-V$
colors vary by $\approx0.5$ mags at fixed $Z$, while $FUV-V$ varies by
$>1$ mag.  This phenomenon is known as the `second--parameter' effect,
and has bee known for decades \citep[see the review in][]{Catelan09}.
Parameters that might be responsible for the additional variation at
fixed $Z$ include age, helium content, mass--loss, stellar rotation,
and magnetic fields.  The particular solution is not as important as
the observed variation.  It is difficult to imagine constraining an
SPS model in the UV given these variations, and it will therefore
continue to be very challenging to extract useful information from the
UV flux of quiescent galaxies.

The locus of ultraviolet colors of Elliptical galaxies do not
correspond to star clusters of any metallicity in either M31, the MW,
or M87.  The $NUV-V$ colors correspond to metal--rich clusters, while
the $FUV-NUV$ colors are similar to metal--poor clusters, and the
$FUV-V$ colors of the Elliptical sample are significantly redder than
any star cluster.  This conflict may be due in part to the fact that
galaxies are not mono--metallic.  The ultraviolet is quite sensitive
to metallicity, and so accurate modeling of the metallicity
distribution function is essential for these objects.  These and
related issues regarding stellar population modeling of Ellipticals
will be discussed in $\S$\ref{s:cqg}--\ref{s:ex}.

It should also be noted that the star clusters in M87 are
significantly bluer in ultraviolet colors at fixed metallicity
compared to clusters from the MW and M31.  The source of this
difference is not known, although differences in cluster age or helium
abundance are possible explanations \citep{Sohn06, Kaviraj07b}.
Nevertheless, it is clear that even if models can be constructed that
match the MW cluster data, such models may not be applicable to other
systems.

Several conclusions can be drawn from the comparisons between models
and data.  First, a 10\% blue HB fraction at all metallicities
produces ultraviolet colors that are comparable to post--AGB stars,
demonstrating the degeneracy between these two evolutionary phases
when considering UV colors.  The BaSTI isochrones and the Padova
isochrones with variable blue HB fraction are both able to fit a
majority of the metallicity--dependent ultraviolet data.  At
metallicities ${\rm log}(Z/Z_\Sol)>-0.7$ these models fail to
reproduce the observed $FUV$ fluxes, whereas models with 100\% blue HB
fractions can reproduce this data.

When modeling the ultraviolet flux of galaxies, a flexible treatment
of hot evolved stars is clearly warranted given our inadequate
theoretical understanding of the observed ultraviolet colors of
clusters, and given the considerable scatter in observed ultraviolet
data at fixed metallicity.

\begin{figure}[!t]
\begin{center}
\plotone{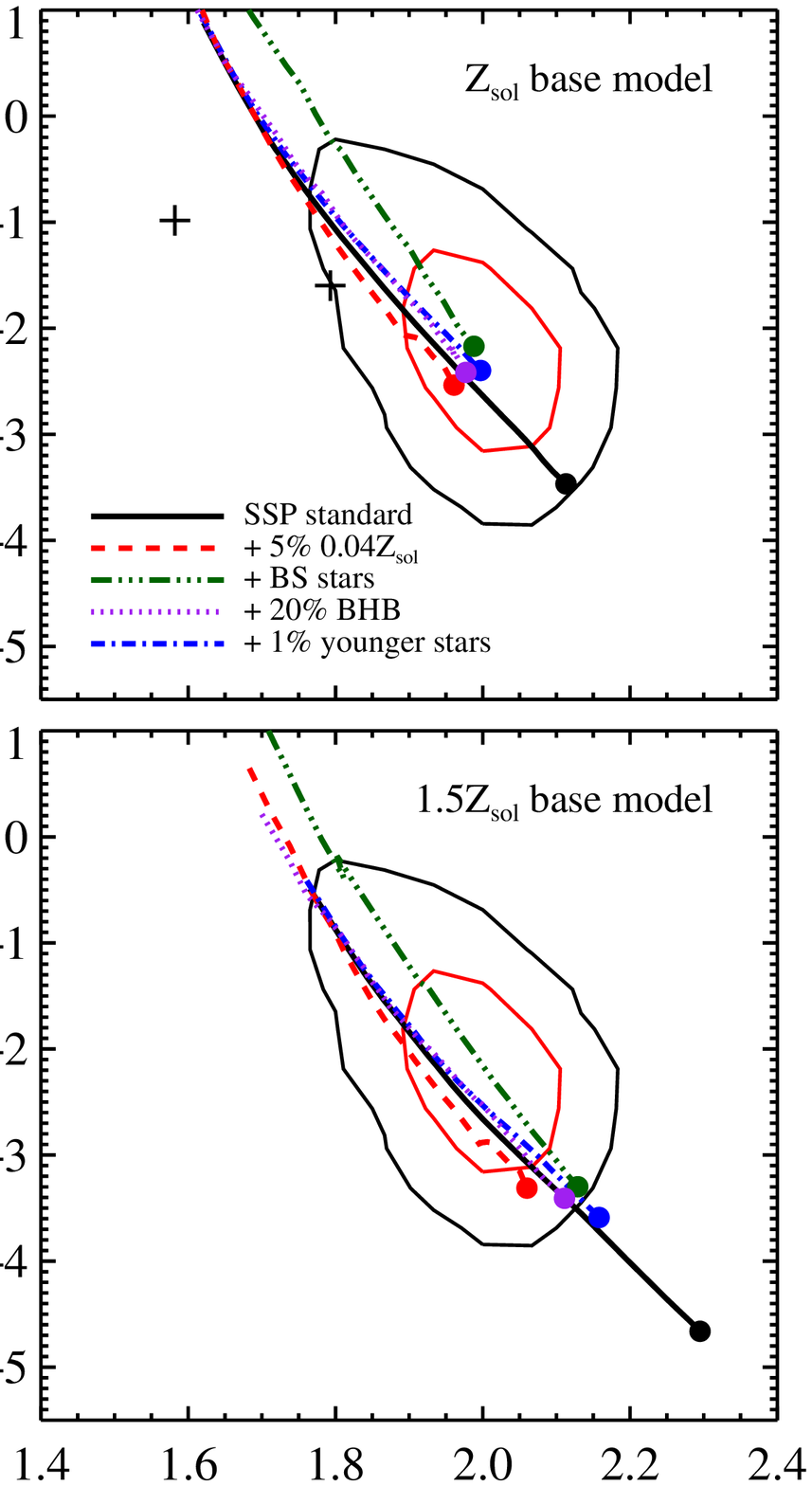}
\end{center}
\vspace{.5cm}
\caption{Comparison of the spectral indices $D_n4000$ and H$\delta_A$
  between SDSS galaxies with $\sigma>200\kms$ ({\it contours}) and
  various models constructed with FSPS adopting the Padova isochrones
  (as in previous figures).  Model spectra are Doppler broadened to
  $\sigma=200\kms$ and span the age range $2<t<13$ Gyr, with the
  oldest point marked by a solid symbol.  Blue HB (BHB) stars, young
  stars (with an age of 1 Gyr), metal--poor stars, and BS stars (with
  $S_{\rm BS}=2$) are considered as additions on top of a base
  dust--free SSP of either solar or super--solar metallicity.  The
  locations of two $\sim Z_\Sol$ MW bulge star clusters are included
  ({\it crosses}) from the integrated spectra of \citet{Schiavon05}.
  The MW star cluster spectra were Doppler broadened to
  $\sigma=200\kms$ for comparison to the galaxy data. }
\label{fig:lick3}
\vspace{0.5cm}
\end{figure}

\begin{figure}[!t]
\begin{center}
\plotone{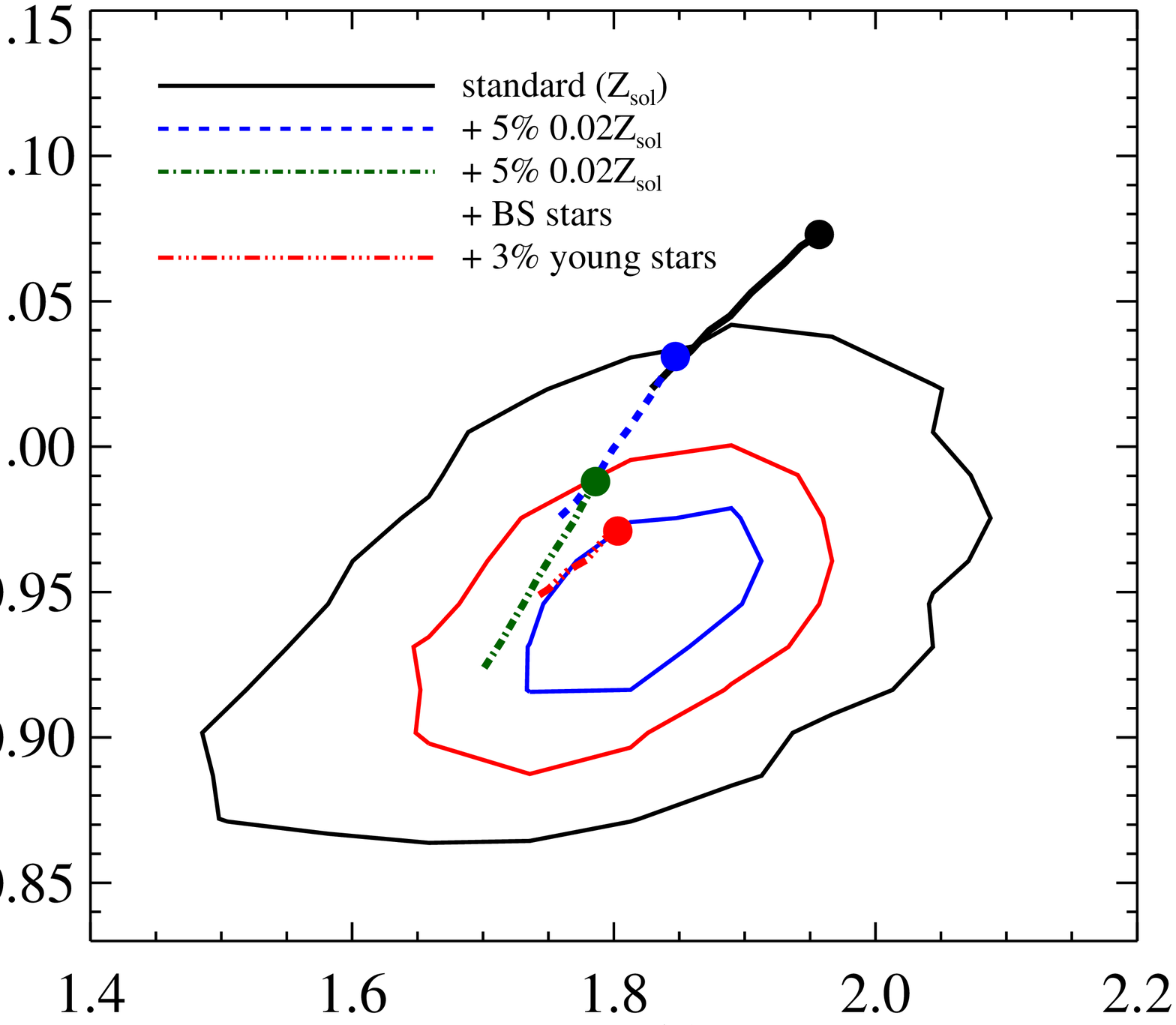}
\end{center}
\vspace{.5cm}
\caption{Comparison of the observed $ugr$ colors of luminous red
  sequence galaxies ({\it contours}) to FSPS models.  All models are
  dust--free SSPs at $Z=Z_\Sol$ for $5<t<13$ Gyr, with symbols marking
  the location of $t=13$ Gyr.  The standard model is compared to
  models where metal--poor stars, BS stars (with $S_{\rm BS}=2$), and
  young stars with an age of 1.0 Gyr, are added on top of the standard
  model in the proportions noted in the legend.}
\label{fig:grcol}
\vspace{0.5cm}
\end{figure}

\subsection{Constraints from quiescent galaxies}
\label{s:cqg}

We now consider constraints on SPS models provided by massive red
sequence, (i.e., quiescent), galaxies.  Such galaxies are thought to
be composed primarily of old ($t>10$ Gyr), metal--rich stars.

\vspace{0.4cm}
\subsubsection{Colors}

Figure \ref{fig:redcol} compares galaxies to models in color--color
space.  We include the FSPS model predictions with the modified Padova
isochrones in addition to the BC03 and M05 models.

All models predict similar $ugr$ colors.  One clearly sees the almost
perfect degeneracy between age and metallicity in this color--color
space.  It is also clear that all models predict $ugr$ colors that are
substantially redder than the data, assuming that such galaxies are
composed of old, metal--rich stars \citep[e.g.,][]{Worthey92,
  Trager00, Thomas05, Graves08}.  The discrepancy between observed and
predicted $g-r$ colors is $\gtrsim0.1$ mags.  Potential remedies to
this discrepancy are explored in $\S$\ref{s:ex}.

The models perform reasonbly well in the near--IR, although there is a
notable disagreement in the $J-H$ color for all models.  In the $r-z$
vs. $H-K$ plane the FSPS and M05 models perform well, while BC03 fits
the data only for solar or sub--solar metallicities.  The BC03 and M05
models predict strong variations in $Y-K$ with metallicity, in
contrast to FSPS.  Given the narrow distribution of observed $Y-K$
colors, the BC03 models would predict nearly all red sequence galaxies
to be solar metallicity, while M05 would predict that all were
super--solar.  In contrast, FSPS allows for a range of metallicities
from sub-- to super--solar, in better agreement with detailed analysis
of the spectra of red sequence galaxies \citep[e.g.,][]{Thomas05}.

Perhaps the most important conclusion to be reached from these
color--color comparisons is that no model is able to reproduce the
observed locus of red galaxies for all near--IR colors, and in
addition no model can match the locus of galaxies in the $u-g$
vs. $g-r$ plane.

It should also be noted that within the context of one model, optical
and near--IR color combinations, in particular $g-r$ vs. $Y-K$, are in
principle capable of breaking the age--metallicity degeneracy.  In
practice it is not currently possible to use such colors to break this
degeneracy for the twofold reason that the models cannot predict the
observed locus of galaxies in that space and the models do not compare
well with each other \citep[see also][]{Eminian08, Carter09}.

The red and near--IR colors of these galaxies are determined primarily
by the RGB.  As noted in $\S$\ref{s:metalr}, resolved CMDs do not help
constrain model predictions for integrated red and near--IR colors
because these colors are determined by the brightest and rarest RGB
stars.  Homogeneous, integrated near--IR photometry for a sample of
old metal--rich clusters would be invaluable for constraining models
in this regime.

\subsubsection{Spectral indices}

Figure \ref{fig:lick1} compares spectral indices of SDSS galaxies to
the FSPS model constructed with the Padova isochrones and the Miles
empirical stellar library. The data are split according to their
velocity dispersions.  All models in this section are broadened to
$\sigma=200\kms$ in order to compare to the more massive data bin.
Model predictions are shown for dust--free SSPs at $Z=Z_\Sol$ and
$Z=1.5Z_\Sol$ for ages $2<t<13$ Gyr.

The index $D_n4000$, which measures the strength of the CaII lines and
the balmer decrement, is often taken to be sensitive primarily to age,
and this is borne out in the figure.  The majority of the other
indices are sensitive to a combination of age and metallicity.  The
index C4668, which is a blend of many metallic lines including Fe, Ti,
Cr, Mg, Ni, and C$_2$ \citep{Worthey94b}, is the most sensitive to
metallicity of the indices shown, at fixed D$_n4000$.  The index
[MgFe]' was defined by \citet{Thomas03} in order to be insensitive to
$\alpha-$enhancement.  The G4300 index is sensitive to both CH and FeI
\citep{Worthey94b}.  H$\delta_A$ is often used to constrain the
fraction of recent star formation because of its sensitivity to A
stars \citep[e.g.,][]{Kauffmann03a}, and so the reliability of SPS
models in the H$\delta_A-{\rm D}_n4000$ plane is particularly
important.  For old populations, the interpretation of H$\delta_A$ is
complicated by the presence of nearby absorption due to the CN
molecule.  This implies that the H$\delta_A$ index can be sensitive to
non--solar abundance ratios, depending on the precise definition of
the index \citep{Prochaska07}.  The trend is such that increasing
$\alpha-$enhancement will result in a weaker measured H$\delta_A$
index because the flux redward of the line will be depressed due to
increased CN absorption.

Figure \ref{fig:lick2} compares several SPS models to the
$\sigma>200\kms$ SDSS sample.  The BC03 model predictions are compared
to the FSPS models using both the modified Padova and BaSTI isochrones
with the Miles empirical stellar library.  The M05 model is not
included because the spectral library in that model (The Kurucz
library) is not of sufficient resolution to measure these indices.

The BaSTI and Padova isochrones have very similar main sequence
turnoff points at old ages (see Figure \ref{fig:cmd}).  The primary
difference between these isochrones is that the BaSTI RGB is
approximately 100K hotter than the Padova RGB.  Blueward of
$\lambda=5000$\AA\, the main sequence and sub--giant branch dominate
the flux, and so one would expect very little differences in spectral
indices between these two isochrone sets in the blue, for a fixed set
of empirical stellar spectra.  Indeed, in Figure \ref{fig:lick2} there
are almost no differences between the two isochrone sets for those
indices that are measured at $\lambda\lesssim5000$\AA.  The only index
that shows appreciable differences is TiO$_2$, which is measured at
$\lambda\approx6230$\AA, where the RGB contributes more to the
integrated flux than the main sequence turnoff point.

The BC03 and FSPS models compare favorably for the D$_n4000$, [MgFe]',
G4300, and TiO$_2$ indices (except for the $Z=1.5Z_\Sol$ models for
the latter index).  The $Z=1.5Z_\Sol$ FSPS model predictions for C4668
compare more favorably with the data than the BC03 model of comparable
metallicity, while the converse is true for the Fe4531 index.  The
differences between the BC03 and FSPS models are most dramatic in the
H$\delta_A-{\rm D}_n4000$ plane, where the BC03 H$\delta_A$ EWs are
larger than the FSPS predictions by $\approx1.3$\AA, and are in
significant discord with the data.  This discrepancy has been noted
previously by \citet{Wild07}, and is most likely due to deficiencies
in the STELIB empirical stellar library used by BC03.

One of the most striking results of these comparisons is that all
models tend to over--predict the D$_n4000$ strength and under--predict
the H$\delta_A$ strength, especially for $Z=1.5Z_\Sol$.  As can be
gleaned from the symbols in the upper left panel of Figure
\ref{fig:lick1}, at $Z=1.5Z_\Sol$ the models reproduce the observed
D$_n4000$ locus of massive galaxies for ages $3-6$ Gyr, while for
$Z=Z_\Sol$ the observations are reproduced for somewhat older ages of
$6-10$ Gyr.  From a detailed analysis utilizing multiple spectral
indices, \citet{Thomas05} find most massive ellipticals to have solar
to super--solar metallicities and ages $>10$ Gyr \citep[see also,
e.g.,][]{Bower92, Bender96, Ziegler99, Graves08}. In light of these
results, an explanation is required for the discrepancy between
observed and predicted D$_n4000$ and H$\delta_A$ values.  Possible
explanations are discussed in the following section.

Before considering possible explanations we point out that the proper
inclusion of $\alpha-$enhancement will only exacerbate the discrepancy
between models and data in regards to the H$\delta_A$ index.  This
occurs because $\alpha-$enhancement will result in an even weaker
H$\delta_A$ index \citep{Prochaska07}.

\begin{figure*}[!t]
\plotone{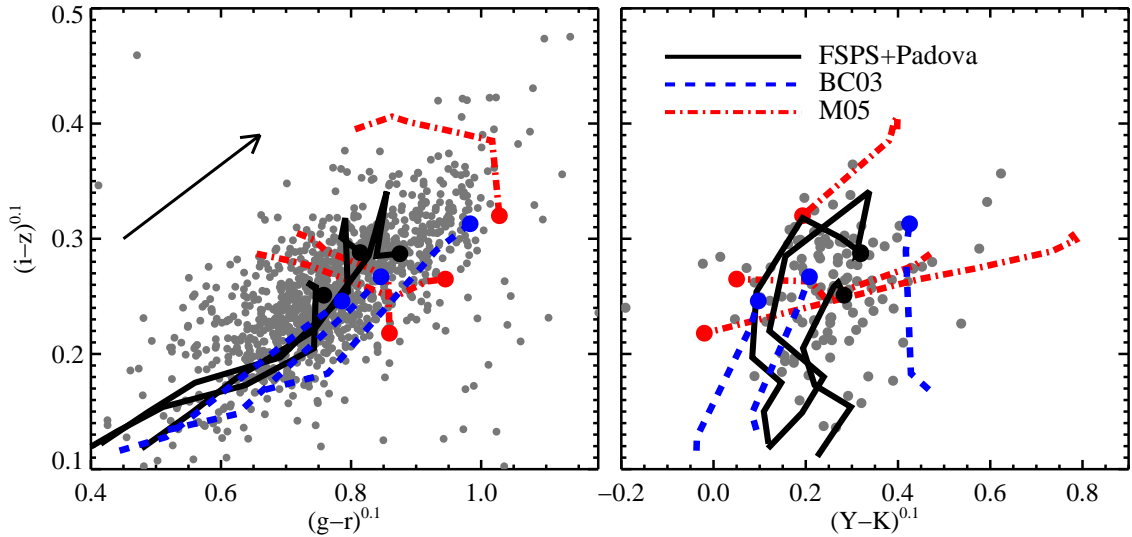}
\vspace{1.cm}
\caption{Color--color diagrams comparing observed K+A
  (post--starburst) galaxies ({\it grey symbols}) to both our FSPS
  model and the models of BC03 and M05.  For BC03 and M05, the colors
  of SSPs are plotted for ages $0.8<t<2.5$ Gyr for three
  metallicities: ${\rm log}(Z/Z_\Sol)\approx-0.2,0.0,0.2$
  (corresponding to three lines for each model).  FSPS models are
  shown for a $\tau-$model SFH with $\tau=0.1$ Gyr, for the same age
  and metallicity ranges as the other models.  The choice of a
  composite SFH for FSPS was chosen for cosmetic reasons only.
  Symbols along the model predictions mark $t=2.5$ Gyr.  In the left
  panel the arrow indicates the change in colors that would occur from
  adding dust with $A_V=0.5$ and a \citet{Calzetti94} reddening curve.
  As discussed in the text, the addition of an underlying old
  population moves the models along the reddening vector, which is
  also along the locus of observed $griz$ colors.}
\label{fig:kacol}
\vspace{0.5cm}
\end{figure*}

\subsection{Resolving the model--data discrepancies for quiescent
  galaxies}
\label{s:ex}

In the previous section two striking disagreements were noted between
SPS models and a sample of massive red sequence galaxies.  Assuming
that such galaxies are composed of `normal' stars\footnote{`Normal'
  here is meant to exclude such stellar phases as blue stragglers and
  hot horizontal branch stars.} of at least solar metallicity and with
ages $>10$ Gyr, all SPS models predict $ugr$ colors that are too red,
D$_n4000$ strengths that are too large, and H$\delta_A$ strengths that
are too weak.

There are two possible classes of solutions to this discrepancy.  In
the first class, an appeal is made to deficiencies in either the
stellar atmospheres or stellar evolution calculations.  Within this
class, an appeal to stellar evolution means considering inadequacies
in standard phases such as the main sequence and its turn--off point,
and the RGB.  Solutions of the second class abandon one or more of the
assumptions regarding the composition of such galaxies.  In this
second class appeals are made to minority populations of either young
stars or metal--poor stars, or exotic aspects of stellar evolution
including blue stragglers and mass-loss leading to an extended
horizontal branch.  These classes are conceptually different in that
the former appeals to a modification of the bulk metal--rich old
population while the latter appeals to additions of minority
populations.  We consider both of these classes below.

\subsubsection{Problems with stellar atmospheres and stellar evolution}

We find that the discrepancies are unlikely to be due to the stellar
spectral libraries.  We have considered both the theoretical BaSeL
library and the empirical Miles library and find similar disagreements
in the colors.  The empirical STELIB library used in BC03 yields
similar disagreements between both the $g-r$ color and the D$_n4000$
strengths, further suggesting that the libraries are not at fault.

These results are at odds with \citet{Maraston09} who showed that the
optical spectra and $g-r$ colors of SSPs at $Z_\Sol$ are markedly
different when comparing models constructed using the BaSeL spectral
library and the empirical \citet{Pickles98} library.  The purported
discrepancies caused by the spectral libraries cannot be reproduced
with our FSPS code when using the empirical Miles library.  To test
that this was not due to differences between the Miles and Pickles
library, we also compared spectra generated from the BC03 model using
both the BaSeL and Pickles library and again find no substantial
difference in $g-r$ colors.  We have also compared the spectra of
individual stars from the Pickles, Miles, and BaSeL libraries for the
same stellar parameters discussed by Maraston et al. and again cannot
reproduce their results (i.e., their Figure 4).

We also point out that in the context of the explanation proposed by
Maraston et al., the BaSeL library should provide a better fit to the
data when the colors are k--corrected to $z=0.0$ (as opposed to
$z=0.1$), because their discrepancy arises in a region of the spectrum
sampled by the $r-$band redshifted to $z=0.1$.  However, when we
compare the colors of $z\lesssim0.03$ red sequence galaxies we find
that the model discrepancies remain.

The discrepancy is also unlikely to be due to unresolved issues with
canonical phases of stellar evolution such as the main sequence and
RGB, because the aspects of stellar evolution relevant for $ugr$
colors are adequately described by the models for old, metal--rich
star clusters.  This claim is based on two observations.  First, from
Figure \ref{fig:hZ} it is clear that FSPS can reproduce the main
sequence turn--off point and base of the RGB of old, metal--rich star
clusters.  This is especially the case for $UBV$ colors of the
clusters NGC 6791 and NGC 188.  To the extent that there are residual
discrepancies, the models tend to be somewhat too blue compared to the
data, which is in the opposite sense of the discrepancies noted for
galaxies.  Second, in Figure \ref{fig:CMDf} it was shown that for blue
colors such as $B-V$, one need not probe the brightest portions of the
RGB in order to achieve convergence in the integrated color (the
results for $g-r$ are very similar to $B-V$).  Therefore, integrated
$g-r$ colors of old metal--rich clusters should be accurate because
the main sequence turn--off and the base of the RGB are relatively
well--fit by the model in the CMD.

In sum, we believe that the discrepancies between model and data $ugr$
colors and spectral indices are not a reflection of the limitations of
either stellar atmospheres or canonical aspects of stellar evolution.

\subsubsection{Minority populations in predominantly old metal--rich
  galaxies}

There are two remaining possible solutions to the disagreement between
models and data for massive red sequence galaxies, which we lump into
one class because they share the common property of being minority
additions to the dominant, old metal--rich population.  The first
solution consists of modifications to the SSPs.  This solution
includes the addition of BS or blue HB stars.

The second solution to these disagreements lies not in trying to fix
the SSPs but rather in relaxing the implicit assumption that these
massive red galaxies are composed of mono--metallic, coeval stars.
Indeed, estimates of the metallicities of stars in our own and nearby
galaxies, either from direct spectroscopy or via the colors of giant
branch stars, always yield a metallicity distribution function that is
approximately lognormal with a full width of $\approx1$ dex
\citep{Rich90, Grillmair96, Harris99, Zoccali03, Worthey05,
  Sarajedini05, Koch06}.  Models for the chemical evolution of
galaxies also generically predict that galaxies should contain stars
of a range of metallicities, including a non--trivial fraction of
metal--poor stars \citep{Searle72, Pagel97}.

Extreme/blue HBs exist in the solar and super--solar metallicity star
clusters NGC 188 \citep{Landsman98} and NGC 6791 \citep{Landsman98,
  Kalirai07}, in addition to moderately sub--solar clusters
\citep{Rich97}.  Resolved UV photometry in the nucleus of M32 has
shown that $\sim5$\% of HB stars are hot \citep{Brown00}.  Moreover,
many metal--rich clusters contain copious BS populations
\citep[e.g.,][]{Landsman98, Zoccali01, Kalirai07}.  These observations
suggest that it would be prudent to consider models that allow for
blue/extended HBs and BS stars to be associated with metal--rich
populations.  In our treatment we are agnostic as to which population
(metal--rich or metal--poor) these hot stars are associated with.

These various effects are explored in Figures \ref{fig:lick3} and
\ref{fig:grcol}.  In these figures we explore the impact of various
modifications to the base dust--free SSP models.  Modifications
include: 1) a small fraction ($1-3$\%) of young ($t=1.0$ Gyr) stars;
2) a population of blue HB stars ($f_{\rm BHB}=0.2$); 3) a population
of BS stars ($S_{\rm BS}=2$); 4) a minority population (5\%) of
metal--poor stars.

All of these modifications move the base SSP models in the correct
qualitative sense; that is they result in lower values of D$_n4000$
and larger H$\delta_A$ strengths, and bluer $ugr$ colors.  These
trends result from the fact that each modification is adding a
population of relatively hot stars, whether in the form of a hotter
main sequence turnoff (as happens when metal--poor stars are added),
hot evolved stars including BS and blue HB stars, or hot young stars.
Hotter stars have lower values of D$_n4000$, larger balmer absorption,
and bluer continua.  

\subsubsection{Summary and outlook}

The discrepancies discussed above between SSP predictions and massive
Elliptical galaxies have been known, in one form or another, for
decades \citep[e.g.,][]{Faber77, Oconnell80, Gunn81, Rose85,
  Schweizer90, Schweizer92, Worthey96, Maraston00, Caldwell03,
  Trager05, Schiavon07}.  Our contribution to this issue has been
twofold: 1) to highlight the connection between the indices and
broadband $ugr$ colors, in the sense that whatever hot star population
is adopted, it should alleviate the discrepancies seen in both indices
and colors; 2) to remind the user of SPS models that reliable results
concerning the ages and metallicities of massive Ellipticals cannot be
attained unless these minority hot star populations are carefully
considered.

Many previous studies have considered one or more of the minority
populations considered herein as a remedy to the model discrepancies.
\citet{Schiavon07} disfavors appeals to blue HB stars based on the
consideration of higher--order balmer lines in Ellipticals.  However,
this conclusion rests on the assumption that the temperature
distribution along the HB in other galaxies is similar to that seen in
the MW.  While this may seem to be a relatively benign assumption,
recent UV observations of clusters in M87 cast doubt on this
\citep{Sohn06}.  If the M87 result holds up to further scrutiny, it
will mean that insights from the MW star cluster population cannot be
readily applied to other galaxies.  This would be a dramatic setback
to our attempts at understanding the stellar populations in other
galaxies, and would have potentially profound implications for our
understanding of stellar evolution.

The addition of metal--poor stars is rarely considered without also
considering the addition of blue HB stars.  In this work we have
separated these two populations, since the observed relation between
metallicity and HB morphology in the MW is complex and displays
considerable scatter \citep{Piotto02}.  It should be remembered that
in every galaxy where metallicity distribution functions have been
measured, there exist metal--poor stars (log$(Z/Z_\Sol)\lesssim-1$) in
sufficient numbers \citep[$\approx1-5$\%; e.g.,][]{Harris99,
  Zoccali03, Sarajedini05} to effect the integrated blue/UV flux.
Given this observational fact, metal--poor stars {\it must} be
included in the models; the only question that remains is in what
proportion.

It is something of a disappointment that the identification of the hot
star population in Ellipticals remains unresolved more than three
decades after it was first discussed.  Despite enormous advances in
both the SPS models and the quality of the data, the fundamental
problem remains that these various hot star populations occupy
essentially the same (or similar) ranges in $T_{\rm eff}$.  We briefly
comment on several possible future directions that might help separate
the relative influence of these various populations.

If young stars or metal--poor stars are the dominant contributor, then
because their metallicities will differ from the bulk population,
metallicities measured in the blue/UV will differ from metallicities
measured in the red.  Models will have to be constructed to determine
whether or not this effect is measurable.  If it is then this could be
a promising means of discriminating between the various hot star
candidates.

If young stars are the culprit, then lookback studies of Ellipticals
will uncover Ellipticals with younger and younger minority
populations.  Such a study would suffer from the typical difficulties
in lookback studies, namely that precise descendant--progenitor
relations are difficult to construct.  However, with new and upcoming
wide--field spectroscopic surveys we believe that the difficulties
associated with lookback studies can be mitigated.  Lookback studies
might also be able isolate the importance of blue HB stars
\citep{Atlee09}.

A final, somewhat more speculative possibility, is the use of SBFs in
the blue to discriminate between these various populations.  While
these populations occupy approximately the same range of $T_{\rm
  eff}$, they all occupy rather different ranges in $L_{\rm bol}$, and
they will thus produce different SBF magnitudes in the blue.  We are
currently constructing models aimed at determining whether or not such
an effect is observable with current facilities.

\subsection{Constraints from post--starburst galaxies}
\label{s:cpsg}

K+A, or post--starburst, galaxies offer a unique constraint on
intermediate age populations.  The prominent A star spectra in these
galaxies without any accompanying H$\alpha$ emission (i.e., no {\it
  current} star formation), indicates that they are dominated by stars
with ages $\approx 1-2$ Gyr.  Such galaxies can therefore provide
potentially strong constraints on the importance of TP--AGB stars,
since such stars peak in prominence at ages $\approx 1-2$ Gyr.

Figure \ref{fig:kacol} compares the colors of K+A galaxies defined by
\citet{Quintero04} to the colors of dust--free SSPs at ages
$0.8<t<2.5$ Gyr for metallicities ${\rm log}(Z/Z_\Sol)\approx
-0.2,0.0,0.2$.  Model predictions from FSPS (using the Padova
isochrones), M05, and BC03 are shown.  The FSPS model predictions are
shown for a $\tau-$model SFH (SFR$\propto e^{-t/\tau}$) with
$\tau=0.1$ Gyr for cosmetic reasons only.  Recall that the TP--AGB
phase in the Padova isochrones was modified in order to produce
acceptable agreement with both the MC star clusters and these
post--starburst galaxies (although the modifications are slight for
intermediate ages at solar metallicity; see $\S$\ref{s:mod}).  While
not shown, FSPS predictions using the BaSTI isochrones are similar to
the Padova isochrones.  A reddening vector is shown to indicate the
shift in colors expected for a uniform screen of dust with optical
depth $A_V=0.5$ and a \citet{Calzetti94} attenuation law.

The ratio of A to K stars in the observed spectra decreases with
increasing $g-r$ and $i-z$ colors, indicating that the sequence in the
left panel is primarily a sequence in age, and not dust nor
metallicity.  This observational fact will be a significant
discriminator amongst the SPS models considered herein.

The FSPS model predictions encompass the range of observed $i-z$,
$g-r$, and $Y-K$ colors.  The dominant variable driving the change in
model colors is age, in agreement with the data.  The BC03 model
predictions are somewhat too blue in $i-z$, although we believe that
the agreement is acceptable given the fact that post--starburst
galaxies are composite populations whereas we have modeled them as
single populations (see below).

In contrast to the FSPS and BC03 models, the M05 model performs poorly
in $g-r$ and $i-z$ colors.  In particular, model sequences of
increasing age produce trends in these colors that are orthogonal to
the observed trends with age.  In the context of the M05 model the
increasing $g-r$ and $i-z$ colors would be interpreted as a sequence
of either increasing dust content or increasing metallicity.  This
interpretation would be in contrast with the observed variation in the
spectral properties.  It is also difficult to explain the bluer half
of the observed K+A galaxies in $griz$ colors unless unrealistically
low metallicities are considered.  Appealing to younger ages does not
remedy this either, as M05 model predictions at younger ages ($t<0.8$
Gyr) continues to diverge from the locus of observed colors.

As discussed in previous sections, it is difficult to reliably
calibrate SPS models against galaxies because galaxies contain stars
of a range of metallicities and ages, and the effects of dust cannot
be ignored.  Here we have been considering a dust--free,
mono--metallic, coeval population model.  For the comparisons in this
section, these complications turn out to have minimal consequences
regarding the model comparisons.  For example, as shown in Figure
\ref{fig:kacol}, the dust reddening vector lies parallel to the locus
of data, and so dust effects cannot move the models onto or off of the
observed locus, but only along the locus.  Since our conclusions do
not rely on the detailed location of the models along the observed
locus, our results are insensitive to dust effects.

We have also considered the effects of adding an old population on top
of the intermediate age, post--starburst population.  Even in the
extreme case where 90\% of the mass is contained in an old population
and only 10\% spans the intermediate age range of $0.8<t<2.5$ Gyr, the
red end of the model predictions only shift slightly toward redder
colors.  The primary effect of an addition of an old population is a
narrowing of the range in color of the model predictions.  In any
event, the addition of old stars has an effect similar to dust
attenuation --- namely, the models shift along the observed locus.  We
conclude that, in the case of post--starburst galaxies and for the
colors we have considered, composite population effects do not impact
our conclusions.

%-----------------------------------------------------------------

\section{Model evaluation}
\label{s:disc}

In the previous section we explored the constraints on several SPS
models provided by star cluster data in the Local Group and M87 and by
post--starburst and massive red sequence galaxies.  We now summarize
these results and provide a general evaluation of the models.

Star clusters in the MCs provide valuable constraints on models at
intermediate age and sub--solar metallicities.  The M05 model performs
poorly when compared to the extant data, while the BC03 model performs
better, although the BC03 model is somewhat too blue in the near--IR.
In M05, the colors are too red and the age--dependence is incorrect.
The incorrect age--dependence in M05 cannot be attributed to the
details of MC cluster age estimation.  Our own FSPS model with Padova
isochrones is able to achieve excellent agreement with the data only
after we modify the TP--AGB phase of the input Padova isochrones.  The
BaSTI isochrones in FSPS fare well, although less--so in the near--IR,
which plausibly reflects their simplistic TP--AGB treatment, and, at
young ages, which may be due to our use of the BaSTI isochrones that
suppresses convective overshooting.  Similar conclusions are reached
regarding the FSPS model when considering constraints afforded by SBFs
of MC star clusters.

The MW and M31 provide a large sample of old, metal--poor star
clusters, but only a small handful of old metal--rich clusters.  The
FSPS (using both the Padova and BaSTI isochrones), BC03, and M05
models all make similar predictions for the $UBVRIJHK$ photometry of
old metal--poor systems.  The models compare favorably in $UBVRI$
colors once uncertainties in the filter transmission curves are
considered.

In the near--IR the models perform less well, where for example $V-K$
colors are too blue by $\approx0.2$ mags at log$(Z/Z_\Sol)>-1.0$.  The
origin of this discrepancy is unclear, but may either be due to an RGB
that is too hot (lowering the RGB by 200K would alleviate this
tension) or stellar libraries that are inadequate in the near--IR.  At
higher metallicity, multi--color CMDs in $UBVIK$ colors are well
reproduced by the FSPS model in regards to the location of the
turn--off point, sub--giant branch, and the lower portion of the RGB.
We demonstrate that agreement in CMD--space between data and models
does not imply that the integrated colors are adequately constrained
because the integrated colors in the red and near--IR are dominated by
the brightest stars in the cluster.  These stars are rarely included
in CMD studies because they are rare and often saturate the
photometry.  In any event, even if these stars are considered, owing
to their rarity it is challenging to reliably calibrate models in this
regime in CMD--space.  It is for this reason that we focus our
calibrations on integrated cluster properties, rather than resolved
CMDs.

The spectral indices D$_n4000$ and H$\delta_A$ of MW star clusters are
not adequately reproduced by either the FSPS or BC03 model at high and
low metallicities, respectively (the M05 model does not contain
sufficient spectral resolution to make a robust prediction).  These
inadequacies are removed if BS or hot HB stars are added.

In the ultraviolet, the FSPS model can reproduce the observed trends
because the relevant phases --- post--AGB and HB --- are handled
flexibly and so can be tuned to match the data.  The BC03 and M05
models generally fare less well, either because of the omission of
blue HB stars (BC03) or post--AGB stars (M05).

Turning to galaxies, all models predict $ugr$ colors of red sequence
galaxies that are too red and the FSPS and BC03 models predict
D$_n4000$ strengths that are too strong (again, the M05 model does not
contain the requisite resolution), under the assumption that such
galaxies are composed solely of old stars with $Z\geq Z_\Sol$.  These
issues are likely closely connected to the model failures at
reproducing the spectral features of metal--rich MW star clusters.
These discrepancies can be removed by adding a minority population of
hot stars, coming either from metal--poor turn--off stars, hot HB
stars, BS stars, or young ($\sim1$ Gyr) stars.

The FSPS and BC03 models are able to reproduce the optical and
near--IR colors of post--starburst galaxies, while the M05 model
performs less well.  These galaxies likely have a preponderance of
TP--AGB stars, and so they provide valuable constraints on this phase.

In summary, our FSPS model is able to reproduce the greatest range of
observations, due in large part to the flexibility of the model.  BC03
fares somewhat less well, and M05 performs worst of the three.  Even
where calibrating data is available, significant discrepancies remain.
Important regions of parameter space do not contain sufficient
calibrating data, and so the models are even less reliable there.
Finally, when modeling predominantly old, metal--rich populations,
care must be taken to include minority populations of hot stars.

\subsection{Comments on other SPS models}

In this section we comment on a number of other SPS models in the
literature.  The main purpose of the following discussion is to
highlight the fact that many SPS models rely on the same ingredients
as the models considered in detail herein, and therefore conclusions
reached in earlier sections should be applicable to other models as
well.

The most popular SPS models not considered herein are Starburst99
\citep{Leitherer99, Vazquez05} and Pegase \citep{Fioc97, LeBorgne04}.
In its most current version, Starburst99 combines the Geneva models
for massive stars and the Padova isochrones for intermediate and low
mass stars.  Starburst99 employs the TP--AGB models of
\citet{Vassiliadis93}, which are the same models used in BC03.  The
BaSeL stellar library is used to convert the models into observable
SEDs.  For hot stars, the theoretical models of \citet{Smith02} are
used.

For stellar populations not dominated by massive stars, Starburst99
will produce results similar to BC03.  The treatment of nebular
emission and massive stars, including Wolf--Rayet stars, is more
sophisticated than either FSPS, M05, or BC03, and should therefore
continue to be the model of choice when emission from massive stars is
important.  The recent Popstar model \citep{Molla09} also contains a
sophisticated treatment of massive stars, including the theoretical
atmosphere models for O and Wolf--Rayet stars of \citet{Smith02}, and
will therefore produce results similar to Starburst99.

Pegase utilizes the Padova isochrones and couples these to two
different stellar spectral libraries: BaSeL, and ELODIE
\citep{Prugniel01}.  ELODIE is an empirical library of $\approx1500$
stars.  In \citet{LeBorgne04}, various spectral indices are computed
with Pegase and compared to observations.  The comparisons focus on a
small sample of indices (including H$\beta$, Fe5270, Mgb, and Fe5335)
for Elliptical galaxies and MW star clusters.  It would be fruitful to
consider more comprehensive calibrating comparisons for the Pegase
model, such as is considered herein.  In any event, the underlying
isochrones are similar to those used in BC03, so the predictions of
Pegase should be similar to BC03.

The Vazdekis models are described in \citet{Vazdekis99},
\citet{Vazdekis03}, and Vazdekis et al. (in prep).  They utilize the
Padova isochrones coupled to the empirical Miles stellar library.  For
old populations this model should thus be essentially identical to the
FSPS predictions when the unmodified Padova isochrones are used.  The
Vazdekis model may thus fail to reproduce the observed near--IR colors
of MC star clusters, since, as we demonstrated, the unmodified Padova
isochrones are a poor match to those data.

Significant efforts are currently underway to generate SPS models that
handle $\alpha-$enhancement self--consistently, in the sense that both
the stellar evolution calculations and the atmosphere models have the
same abundance patterns \citep[e.g.,][]{Coelho07, LeeHC09}.  These
models will provide valuable constraints on the relative effects of
$\alpha-$enhancement.  However, most models currently only follow
stellar evolution through the end of the RGB, and so they cannot be
used on their own to build model galaxies \citep[the exception to this
is the BaSTI database;][]{Cordier07, Percival09}.  We anticipate that
the treatment of $\alpha-$enhancement as a variable on par with age
and $Z$ in SPS models will likely become commonplace in the near
future.

From the above discussion it should be clear that the majority of SPS
codes use the Padova isochrones with the BaSeL stellar library.  Our
comparisons and evaluations in previous sections, which rely on these
ingredients, should therefore be applicable to many other SPS models.

%-----------------------------------------------------------------

\section{Summary}
\label{s:sum}

We now summarize our principle results.

\begin{itemize}

\item

  The latest stellar models from the Padova group cannot fit the
  near--IR photometry and surface brightness fluctuations of star
  clusters in the MCs.  With our flexible SPS (FSPS) code we have
  modified the TP--AGB phase in the Padova isochrones to produce
  better fits to the data.  The M05 models also fail to reproduce the
  data, both because the colors are too red and the age--dependence is
  incorrect.  The BC03 and the FSPS model using the BaSTI isochrones
  fare well, although both models are somewhat too blue in the
  near--IR.  A significant limiting factor in more accurate
  calibrations is the difficulty in constructing reliable photometry
  of MC clusters, owing to the fact that TP--AGB stars are luminous
  and rare.

\item

  At low metallicities (log$(Z/Z_\Sol)<-0.5$) and old ages, all models
  predict similar $UBVRIJHK$ colors (the typical variation between
  models is $\approx0.05$ mags).  When compared to MW and M31 star
  clusters, the models are 0.2 mags too blue in $V-K$ at
  log$(Z/Z_\Sol)>-1.0$, and 0.1 mags too red in $J-K$ at all $Z$.  The
  origin of the discrepancies in the near--IR is unclear.  Decreasing
  the temperature of the RGB by $\sim200$K alleviates much of the
  disagreement.  If this modification is isolated to the brightest
  stars, then such a modification could improve agreement with
  integrated photometry without compromising agreement in CMD--space.
  The FSPS model with the BaSTI isochrones performs worst at low
  metallicities and old ages of the models considered (although the
  disagreement is not dramatic), while the FSPS model with the Padova
  isochrones performs well.

  Multi--color CMDs of two metal--rich clusters, NGC 6791 and NGC 188,
  are well--fit by FSPS.  Spectral indices of MW star clusters are
  generally well--fit by both FSPS and BC03, although the models
  predict D$_n4000$ strengths too large and H$\delta_A$ strengths too
  weak compared to two log$(Z/Z_\Sol)\approx0.0$ clusters.

\item

  The ultraviolet photometry of the MW, M87, and M31 star clusters can
  be well--fit by FSPS because FSPS contains flexible treatments of
  the post--AGB and horizontal branch evolutionary phases.  This
  flexibility is essential given the substantial scatter in the
  ultraviolet data at fixed metallicity, and given our inadequate
  theoretical understanding of these phases.  The M05 and BC03 models
  perform less well because of their incomplete treatment of these
  advanced evolutionary phases.

\item

  SPS models are compared to $ugrzYJHK$ photometry of massive red
  sequence galaxies.  The FSPS, BC03, and M05 models fare well in most
  respects, with a few exceptions: all models are too blue in $J-H$;
  the M05 model is somewhat too blue in $Y-K$; all models are far too
  red in $g-r$ and $u-g$.  These conclusions hold under the assumption
  that red sequence galaxies are composed exclusively of old
  metal--rich stars with canonical evolutionary phases.  The
  disagreement in the $ugr$ colors can be alleviated if some
  combination of young ($\sim1$ Gyr) stars, metal--poor stars, and
  blue straggler stars are added in small amounts.

\item

  Optical spectral indices of massive galaxies are generally well--fit
  by the FSPS and BC03 models, with the important exceptions that both
  models predict D$_n4000$ in excess of observations and neither fit
  the observed locus of galaxies in the D$_n4000-$H$\delta_A$ plane.
  These conclusions hold for the same assumptions noted above, and, as
  above, small additions of some combination of young stars,
  metal--poor stars, blue straggler and horizontal branch stars can
  remedy this disagreement.

\item

  The FSPS and BC03 models adequately describe the $grizYK$ colors of
  K+A, or `post--starburst' galaxies, while the M05 model performs
  less well.  Such galaxies contain a large proportion of intermediate
  age ($0.5<t<2$ Gyr) stars and thus provide unique constraints on the
  importance of TP--AGB stars in galaxies.

\end{itemize}

\acknowledgments 

We thank David Hogg for providing the K+A galaxy catalog, Patricia
S{\'a}nchez-Bl{\'a}zquez for providing the Miles library in electronic
format, Rita Gautschy for providing her model results, the BaSTI,
Padova, BaSeL, and MPA/JHU groups for publically releasing their
results and for assistance in their use, and Tony Sohn for help with
interpretation of the ultraviolet star cluster data.  Santi Cassisi,
David Hogg, Claudia Maraston and Ricardo Schiavon are thanked for
comments on an earlier draft.

Funding for the Sloan Digital Sky Survey (SDSS) has been provided by
the Alfred P. Sloan Foundation, the Participating Institutions, the
National Aeronautics and Space Administration, the National Science
Foundation, the U.S. Department of Energy, the Japanese
Monbukagakusho, and the Max Planck Society. The SDSS Web site is
http://www.sdss.org/.

The SDSS is managed by the Astrophysical Research Consortium (ARC) for
the Participating Institutions. The Participating Institutions are The
University of Chicago, Fermilab, the Institute for Advanced Study, the
Japan Participation Group, The Johns Hopkins University, Los Alamos
National Laboratory, the Max-Planck-Institute for Astronomy (MPIA),
the Max-Planck-Institute for Astrophysics (MPA), New Mexico State
University, University of Pittsburgh, Princeton University, the United
States Naval Observatory, and the University of Washington.

This publication makes use of data products from the Two Micron All
Sky Survey, which is a joint project of the University of
Massachusetts and the Infrared Processing and Analysis
Center/California Institute of Technology, funded by the National
Aeronautics and Space Administration and the National Science
Foundation.

This work made extensive use of the NASA Astrophysics Data System and
of the {\tt astro-ph} preprint archive at {\tt arXiv.org}.

%\bibliography{../master_refs}

\end{document}